\documentclass[%
 aip,
 amsmath,amssymb,
 twocolumn,
 reprint,%
]{revtex4-2}

\usepackage{graphicx}% Include figure files
\usepackage{dcolumn}% Align table columns on decimal point
\usepackage{bm}% bold math
\usepackage{xcolor}
\usepackage{url}

\usepackage{booktabs}
\usepackage{makecell}
\usepackage{enumitem}

\usepackage[T1]{fontenc}
\usepackage{textcomp,gensymb} % for \degree

\usepackage{hyperref} % usually, this should be the last package
\hypersetup{
    colorlinks=true,
    linkcolor=blue,
    filecolor=magenta,      
    urlcolor=cyan,
}
\begin{document}

% \preprint{AIP/123-QED}

\title{Characterization and performance of the Apollon Short-Focal-Area facility following its commissioning at 1 PW level}
% Force line breaks with \\

\author{K. Burdonov} 
\affiliation{LULI - CNRS, CEA, Sorbonne Universit\'e, Ecole Polytechnique, Institut Polytechnique de Paris - F-91128 Palaiseau cedex, France}
\affiliation{Sorbonne Universit\'e, Observatoire de Paris, Universit\'e PSL, CNRS, LERMA, F-75005, Paris, France}
\affiliation{IAP, Russian Academy of Sciences, 603155, Nizhny Novgorod, Russia}

\author{A. Fazzini} 
\affiliation{LULI - CNRS, CEA, Sorbonne Universit\'e, Ecole Polytechnique, Institut Polytechnique de Paris - F-91128 Palaiseau cedex, France}

\author{V. Lelasseux} 
\affiliation{LULI - CNRS, CEA, Sorbonne Universit\'e, Ecole Polytechnique, Institut Polytechnique de Paris - F-91128 Palaiseau cedex, France}

\author{J. Albrecht} 
\affiliation{LULI - CNRS, CEA, Sorbonne Universit\'e, Ecole Polytechnique, Institut Polytechnique de Paris - F-91128 Palaiseau cedex, France}

\author{P. Antici} 
\affiliation{INRS-EMT, 1650, boulevard Lionel-Boulet, J3X 1S2, Varennes (Qu\'ebec), Canada}

\author{Y. Ayoul} 
\affiliation{LULI - CNRS, CEA, Sorbonne Universit\'e, Ecole Polytechnique, Institut Polytechnique de Paris - F-91128 Palaiseau cedex, France}

\author{A. Beluze} 
\affiliation{LULI - CNRS, CEA, Sorbonne Universit\'e, Ecole Polytechnique, Institut Polytechnique de Paris - F-91128 Palaiseau cedex, France}

\author{D. Cavanna} 
\affiliation{LULI - CNRS, CEA, Sorbonne Universit\'e, Ecole Polytechnique, Institut Polytechnique de Paris - F-91128 Palaiseau cedex, France}

\author{T. Ceccotti}
\affiliation{LIDYL, Commissariat \`{a} l'Energie Atomique, DSM/DRECAM, CEA Saclay,
91191 Gif sur Yvette, France}

\author{M. Chabanis} 
\affiliation{LULI - CNRS, CEA, Sorbonne Universit\'e, Ecole Polytechnique, Institut Polytechnique de Paris - F-91128 Palaiseau cedex, France}

\author{A. Chaleil} 
\affiliation{Universit\'{e} Paris-Saclay, CEA, LMCE, 91680 Bruy\`{e}res-le-Ch\^{a}tel, France}

\author{S.N. Chen}
\affiliation{”Horia Hulubei” National Institute for Physics and Nuclear Engineering, 30 Reactorului Street, RO-077125, Bucharest-Magurele, Romania}

\author{Z. Chen} 
\affiliation{INRS-EMT, 1650, boulevard Lionel-Boulet, J3X 1S2, Varennes (Qu\'ebec), Canada}

\author{F. Consoli} 
\affiliation{ENEA, Fusion and Technologies for Nuclear Safety Department, C.R. Frascati, Via E. Fermi 45, 00044 Frascati, Italy}

\author{M. Cuciuc}
\affiliation{”Horia Hulubei” National Institute for Physics and Nuclear Engineering, 30 Reactorului Street, RO-077125, Bucharest-Magurele, Romania}

\author{X. Davoine} 
\affiliation{Universit\'{e} Paris-Saclay, CEA, LMCE, 91680 Bruy\`{e}res-le-Ch\^{a}tel, France}

\author{J.P. Delaneau} 
\affiliation{LULI - CNRS, CEA, Sorbonne Universit\'e, Ecole Polytechnique, Institut Polytechnique de Paris - F-91128 Palaiseau cedex, France}

\author{E. d'Humières} 
\affiliation{University of Bordeaux, Centre Lasers Intenses et Applications, CNRS, CEA, UMR 5107, F-33405 Talence, France}

\author{J-L. Dubois} 
\affiliation{University of Bordeaux, Centre Lasers Intenses et Applications, CNRS, CEA, UMR 5107, F-33405 Talence, France}

\author{C. Evrard} 
\affiliation{LULI - CNRS, CEA, Sorbonne Universit\'e, Ecole Polytechnique, Institut Polytechnique de Paris - F-91128 Palaiseau cedex, France}

\author{E. Filippov}
\affiliation{Joint Institute for High Temperatures, RAS, 125412, Moscow, Russia}
\affiliation{IAP, Russian Academy of Sciences, 603155, Nizhny Novgorod, Russia}

\author{A. Freneaux} 
\affiliation{LULI - CNRS, CEA, Sorbonne Universit\'e, Ecole Polytechnique, Institut Polytechnique de Paris - F-91128 Palaiseau cedex, France}

\author{P. Forestier-Colleoni} 
\affiliation{LULI - CNRS, CEA, Sorbonne Universit\'e, Ecole Polytechnique, Institut Polytechnique de Paris - F-91128 Palaiseau cedex, France}

\author{L. Gremillet} 
\affiliation{Universit\'{e} Paris-Saclay, CEA, LMCE, 91680 Bruy\`{e}res-le-Ch\^{a}tel, France}

\author{V. Horny} 
\affiliation{LULI - CNRS, CEA, Sorbonne Universit\'e, Ecole Polytechnique, Institut Polytechnique de Paris - F-91128 Palaiseau cedex, France}
\affiliation{Universit\'{e} Paris-Saclay, CEA, LMCE, 91680 Bruy\`{e}res-le-Ch\^{a}tel, France}

\author{L. Lancia} 
\affiliation{LULI - CNRS, CEA, Sorbonne Universit\'e, Ecole Polytechnique, Institut Polytechnique de Paris - F-91128 Palaiseau cedex, France}

\author{L.  Lecherbourg} 
\affiliation{Universit\'{e} Paris-Saclay, CEA, LMCE, 91680 Bruy\`{e}res-le-Ch\^{a}tel, France}

\author{N. Lebas} 
\affiliation{LULI - CNRS, CEA, Sorbonne Universit\'e, Ecole Polytechnique, Institut Polytechnique de Paris - F-91128 Palaiseau cedex, France}

\author{A. Leblanc}
\affiliation{Laboratoire d’Optique Appliqu\'ee, Ecole polytechnique – ENSTA – CNRS – Institut Polytechnique de Paris, Palaiseau, France}

\author{W. Ma}
\affiliation{State Key Laboratory of Nuclear Physics and Technology, School of Physics, CAPT, Peking University, Beijing 100871, China}

\author{L. Martin} 
\affiliation{LULI - CNRS, CEA, Sorbonne Universit\'e, Ecole Polytechnique, Institut Polytechnique de Paris - F-91128 Palaiseau cedex, France}

\author{F. Negoita}
\affiliation{”Horia Hulubei” National Institute for Physics and Nuclear Engineering, 30 Reactorului Street, RO-077125, Bucharest-Magurele, Romania}

\author{J-L. Paillard} 
\affiliation{LULI - CNRS, CEA, Sorbonne Universit\'e, Ecole Polytechnique, Institut Polytechnique de Paris - F-91128 Palaiseau cedex, France}

\author{D. Papadopoulos} 
\affiliation{LULI - CNRS, CEA, Sorbonne Universit\'e, Ecole Polytechnique, Institut Polytechnique de Paris - F-91128 Palaiseau cedex, France}

\author{F. Perez} 
\affiliation{LULI - CNRS, CEA, Sorbonne Universit\'e, Ecole Polytechnique, Institut Polytechnique de Paris - F-91128 Palaiseau cedex, France}

\author{S. Pikuz}
\affiliation{Joint Institute for High Temperatures, RAS, 125412, Moscow, Russia}

\author{G. Qi}
\affiliation{State Key Laboratory of Nuclear Physics and Technology, School of Physics, CAPT, Peking University, Beijing 100871, China}

\author{F. Qu\'er\'e}
\affiliation{LIDYL, Commissariat \`{a} l'Energie Atomique, DSM/DRECAM, CEA Saclay,
91191 Gif sur Yvette, France}

\author{L. Ranc}
\affiliation{LULI - CNRS, CEA, Sorbonne Universit\'e, Ecole Polytechnique, Institut Polytechnique de Paris - F-91128 Palaiseau cedex, France}
\affiliation{Laboratoire Charles Fabry, Institut d’Optique Graduate School, CNRS, Universit\'e Paris-Saclay 91127, Palaiseau Cedex, France}

\author{P.-A. S\"oderstrom}
\affiliation{”Horia Hulubei” National Institute for Physics and Nuclear Engineering, 30 Reactorului Street, RO-077125, Bucharest-Magurele, Romania}

\author{M. Scisci\`o} 
\affiliation{ENEA, Fusion and Technologies for Nuclear Safety Department, C.R. Frascati, Via E. Fermi 45, 00044 Frascati, Italy}

\author{S. Sun} 
\affiliation{INRS-EMT, 1650, boulevard Lionel-Boulet, J3X 1S2, Varennes (Qu\'ebec), Canada}

\author{S. Valli\`eres} 
\affiliation{INRS-EMT, 1650, boulevard Lionel-Boulet, J3X 1S2, Varennes (Qu\'ebec), Canada}

\author{P. Wang}
\affiliation{State Key Laboratory of Nuclear Physics and Technology, School of Physics, CAPT, Peking University, Beijing 100871, China}

\author{W. Yao} 
\affiliation{LULI - CNRS, CEA, Sorbonne Universit\'e, Ecole Polytechnique, Institut Polytechnique de Paris - F-91128 Palaiseau cedex, France}
\affiliation{Sorbonne Universit\'e, Observatoire de Paris, Universit\'e PSL, CNRS, LERMA, F-75005, Paris, France}

\author{F. Mathieu} 
\affiliation{LULI - CNRS, CEA, Sorbonne Universit\'e, Ecole Polytechnique, Institut Polytechnique de Paris - F-91128 Palaiseau cedex, France}

\author{P. Audebert} 
\affiliation{LULI - CNRS, CEA, Sorbonne Universit\'e, Ecole Polytechnique, Institut Polytechnique de Paris - F-91128 Palaiseau cedex, France}

\author{J. Fuchs} 
\affiliation{LULI - CNRS, CEA, Sorbonne Universit\'e, Ecole Polytechnique, Institut Polytechnique de Paris - F-91128 Palaiseau cedex, France}

\date{\today}% It is always \today, today,
             %  but any date may be explicitly specified

\begin{abstract}

We present the results of the first commissioning phase of the ``short focal length'' area (SFA) of the Apollon laser facility (located in Saclay, France), which was performed with the first available laser beam (F2), scaled to a nominal power of one petawatt. Under the conditions that were tested, this beam delivered on target pulses of 10 J average energy and 24 fs duration. Several diagnostics were fielded to assess the performance of the facility. The on-target focal spot, its spatial stability, the temporal intensity profile prior to the main pulse, as well as the resulting density gradient formed at the irradiated side of solid targets have been thoroughly characterized, with the goal of helping users design future experiments. Emissions of energetic electrons, ions and electromagnetic radiation were recorded, showing good laser-to-target coupling effiency and an overall performance comparable with that of similar international facilities. This will be followed in 2022 by a further commissioning stage at the multi-petawatt level. 

\end{abstract}

\maketitle

\section{\label{sec:intro}INTRODUCTION}

High-power lasers have become indispensable tools to investigate extreme states of matter subject to ultrastrong electromagnetic fields, enabling a plethora of scientific and technical applications including the generation of unprecedentedly dense beams of energetic particles, the development of ultrashort and/or ultrabright photon sources, or the laboratory reproduction of high-energy astrophysical phenomena \cite{danson_2019, lureau_2020}.

The Apollon laser system, near completion in the Orme des Merisiers campus in Saclay, France, will be among the first multi-petawatt (PW) user facilities worldwide devoted to studying laser-matter interactions at laser intensities exceeding $2 \times 10^{22}\,\rm Wcm^{-2}$. The final goal of the Apollon laser is to generate 10~PW pulses of 150~J energy and 15~fs (FWHM) duration at a repetition rate of 1~shot/minute \cite{Papadopoulos:17, papadopoulos_2016, zou_2015}. In its final configuration, the Apollon system will comprise two high-intensity laser beams: F1, with a maximum power of 10~PW, and F2, with a maximum power of 1~PW, will be simultaneously available to users both in the long-focal (LFA) and short-focal (SFA) areas of the facility. The commissioning of the 1~PW F2 beamline has been completed recently, allowing the first laser-plasma interaction experiments to be conducted in both areas \cite{Papadopoulos:19, Papadopoulos:21}. 

In this paper, we report on the current status of the Apollon laser, and present the results of the first commissioning experiment that took place in the SFA in May 2021, using the F2 beamline. This experiment was devoted to qualifying the potential of this laser for particle (electrons and ions) acceleration and x-ray generation, as well as to characterizing the level of the accompanying electromagnetic pulse (EMP) \cite{Dubois_PRE, Consoli_HPLSE, Consoli_Phtransa}.

\begin{figure}[htp]
    \centering
    \includegraphics[width=0.45\textwidth]{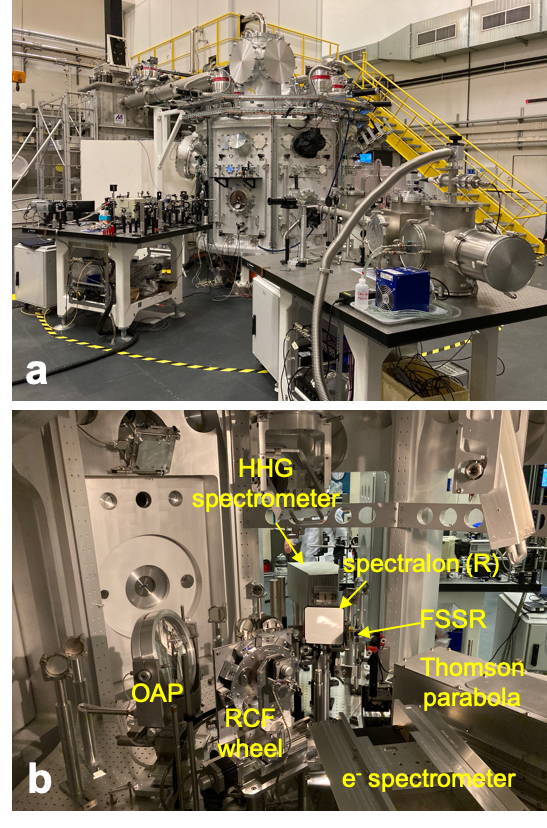}
    \caption{Photographs of the (a) Apollon SFA and of (b) the inside of the vacuum chamber.
    The off-axis parabola (OAP) focusing the laser beam on target is shown. The various diagnostics used during the experiment are also labeled. 
    The high-order harmonic generation (HHG) diagnostic served to measure the spectrum of harmonics of the laser light emitted from the irradiated surface of solid targets. FSSR stands for ``focusing spectrometer with spatial resolution''. RCF stands for ``radiochromic films'' \cite{bolton2014}, i.e., films used to detect the protons stemming from the (non-irradiated) rear side of the solid targets. The label R for the spectralon plate (a Lambertian scatter plate) stands for ``reflection'' since this scatter plate allows the (specularly) reflected fraction of the laser beam to be visualized.
    }
    \label{fig:SFA_photo}
\end{figure}

Figure~\ref{fig:SFA_photo}(a) shows an outside view of the vacuum chamber in the SFA, where the F2 beam is focused to high intensity using a short $f$-number ($F/3$) parabola. Figure~\ref{fig:SFA_photo}(b) displays the inside of the chamber, together with the equipment fielded during the commissioning campaign.

The paper is organised as follows. In Section \ref{sec:status}, we present the parameters of the Apollon PW laser beam. In Section \ref{sec:results}, we give a brief overview of all the diagnostics used for the commissioning campaign, and present the physical parameters that could be retrieved from it. In Section \ref{sec:conclusion}, we discuss the results and draw some conclusions.

\section{\label{sec:status}STATUS OF APOLLON AND ITS F2 BEAM}

\begin{figure}[htp]
    \centering
    \includegraphics[width=0.45\textwidth]{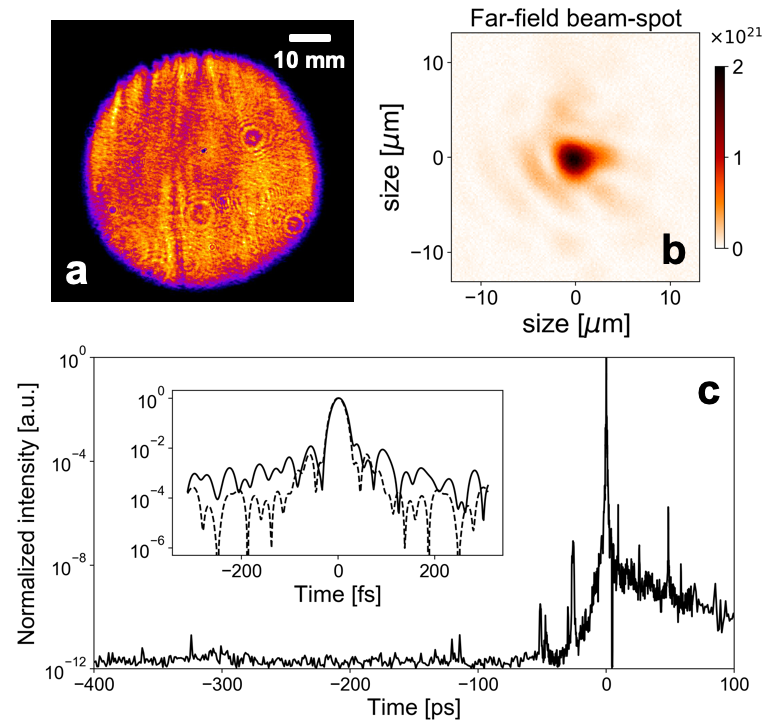}
    \includegraphics[width=0.45\textwidth]{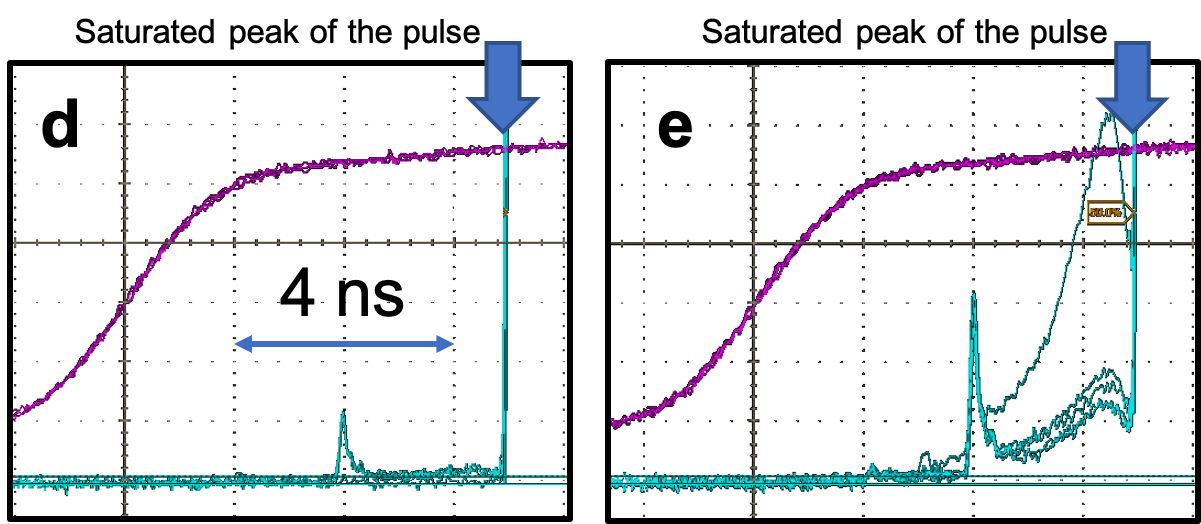}  \caption{(a) Near-field beam profile at the output of the fourth amplification stage (out of five in total) operating at 38 J energy level. (b) Full-energy focal spot as measured in the interaction target after high optical quality neutral attenuation of the beam before compression. (c) Third-order cross correlation intensity contrast measurement at the output of the first Ti:Sapphire amplifier of the Apollon laser; (c inset) typical Wizzler measurement of the compressed 24 fs pulses reaching the interaction chamber (solid line). The Fourier-transform-limited pulse is plotted as a dashed line. (d,e) Oscilloscope screenshots showing the pulse contrast measurements using a fast photodiode at the output of the compressor. Two regimes are illustrated: (d) well-adjusted front-end system and nominal operation of the Apollon laser, (e) non-optimized OPCPA (unstable injection) pump-signal instabilities impact on the end chain contrast. In both cases, the main pulse is saturated on purpose in order to see the weak prepulse ahead of it. Note that the peaked prepulse seen at 3 ns in panel (e) has the same amplitude as in panel (d). In panel (e), the vertical scale of the oscilloscope channel has been enlarged to better visualize the unstable behaviour of the ASE in the last 3-4~ns before the laser peak. The purple trace in panels (d) and (e) is the facility trigger.}
    \label{fig:Laser}
\end{figure}

In the current version of the system, the fifth and final amplification stage (scheduled to provide $>250\,\rm J$ pulses in 2022) is temporarily bypassed. Therefore, a maximum pulse energy of 38~J, from the fourth amplifier, can be delivered with a uniform high-quality flat-top beam (see Fig.~\ref{fig:Laser} (a)). The central laser wavelength is $\lambda_L=815\,\rm nm$, with a spectrum extending over 750–880~nm. For the commissioning, we operated the system at a $\sim 30\,\rm J$ pulse energy level with a typical 1.5\% rms stability over 6 hours of continuous operation. Beam wavefront control was implemented at the output of the amplification section using an adaptive-optics (AO) correction system, consisting of a deformable mirror (DM) with 52 mechanical actuators. A Strehl ratio of $\sim$ 70\%, estimated directly by the focused beam quality, was typically obtained on full-energy shots at the output of the amplification section.
%(the Strehl ratio after focusing is given later).
The 140 mm diameter F2 beam is then directed into the compressor and reaches the target area after about 60 m of free propagation and reflecting off $\sim$ 30 optical interfaces. A residual wavefront error of $\sim$ 1.2 $\lambda$ PtV (principally astigmatism) is corrected by manual adjustment of the DM in the amplification section. The resulting beam quality, once focused on target by the $F/3$ focusing parabola [see Fig.~\ref{fig:SFA_photo}(b)], is illustrated in Fig.~\ref{fig:Laser}(b). 
The focal spot is recorded by an imaging system positioned after the focus, inside the target chamber. In order not to damage that imaging system, the fully amplified beam is attenuated before compression and transport to the target chamber. About 41~\% of the total laser energy is enclosed within a disk of diameter equal to the $2.8\,\rm \mu m$ FWHM. For a further automatized optimization of the end chain beam, a second AO correction loop on the target will be installed and commissioned in 2022.

Temporal compression is performed by means of a folded compressor composed of two gold gratings (1480 l/mm). Pulses of typical 24 fs (FWHM) duration (close to 23.8 fs Fourier-transform limit) are measured using a Wizzler device, see the inset of Fig.~\ref{fig:Laser}(c). The pulse contrast is characterized via different techniques to cover the maximum temporal range. Figure~\ref{fig:Laser}(c) displays a typical third-order cross correlation measurement conducted at the output of the first Ti:Sapphire amplifier, with a a $\sim 7 \times 10^{11}$ dynamic range. A clean pulse pedestal is evidenced up to 400 ps before the main peak. A few prepulses are present closer to the peak (especially at -52 ps and -26 ps) and their nature is under investigation \cite{Ranc:20}. Figures~\ref{fig:Laser}(d,e) show temporal intensity profiles as measured at the output of the compressor using a fast photodiode and calibrated optical densities. These single-shot measurements, carried out at full power using a 10 mm sub-aperture of the beam, have a $\sim 100\,\rm ps$ temporal resolution over a 100~ns timespan. They reveal two salient features. First, an intense prepulse is seen at $\sim 3~\rm ns$ before the main peak. This prepulse, here characterized by an energy contrast of $\sim 10^8$, arises from the on-axis diffusion term of the beam generated by the second amplifier’s Ti:Sapphire crystal, just before the last pass in the amplifier. Taking account of the laser's low spatial coherence and imperfect compressibility, a conservative estimate of the on-target intensity contrast is $\sim 10^9-10^{10}$. Second, under nominal conditions [see Fig.~\ref{fig:Laser}(d)], the amplified spontaneous emission (ASE) noise level is found to lie in the $10^{11}-10^{12}$ range, in agreement with cross-correlation measurements. However, on-shot measurements exhibit sporadic, yet severe pulse-to-pulse instabilities of the ASE contrast, in the last 3-4 ns before the laser peak, 
%A detailed analysis is ongoing, but the cause of this effect is related to the OPCPA stage in the Front End and more specifically timing instabilities/drifts between the pump and the signal pairs.
related to injection instabilities of the front-end (FE) source. These are consequences of optimization procedures of the FE source (performed on a weekly basis) not being strictly applied over the full course of the experimental campaign. This can lead the ASE level in the last 3-4~ns before the main peak to rise by almost 2 orders of magnitude [see Fig.~\ref{fig:Laser}(e)], resulting in failed shots (these making up about 2-3\% of the total number of shots). Careful monitoring and active correction of these instabilities is scheduled for the next experimental campaigns. 

Given the beam transport losses ($\sim 20\%$) and the compression efficiency ($\sim 66\%$), the maximum energy reaching the interaction chamber is estimated to be $\sim 15\,\rm J$. When further including the losses of the uncoated protection silica film of the focusing parabola, the maximum laser energy reaching the target should be of $\sim 10\,\rm J$. From the recorded focal spot shown in Fig.~\ref{fig:Laser}(b), this leads to an on-target peak intensity close to $2 \times 10^{21}\,\rm Wcm^{-2}$.

\begin{figure}[htp]
    \centering
    \includegraphics[width=0.45\textwidth]{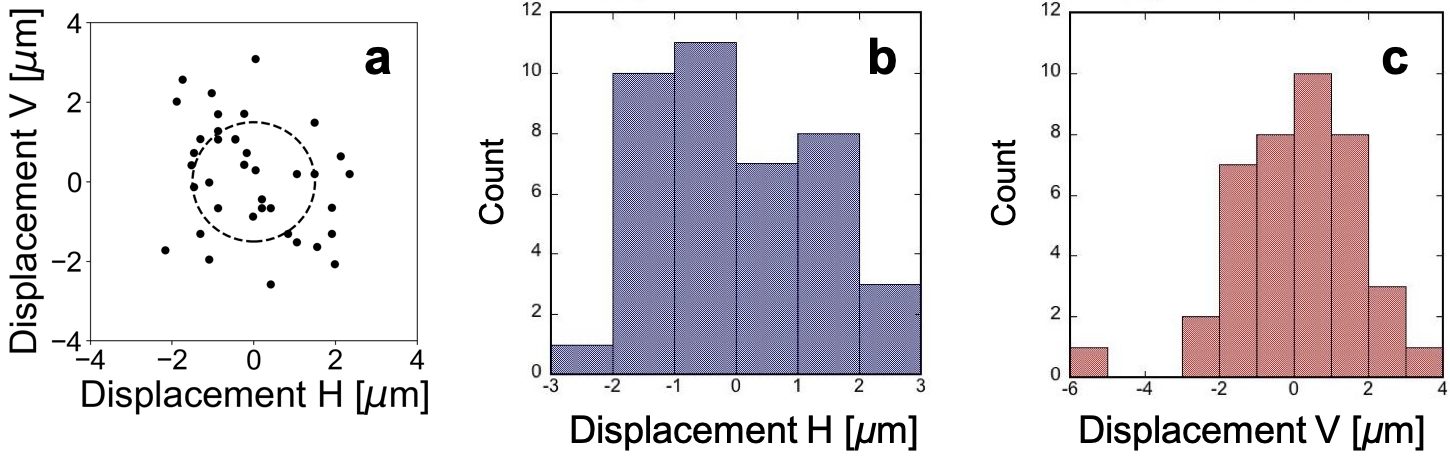}
    \caption{(a) Laser focus beam-spot centroid motion with time; the dashed circle marks the diameter of the central peak of the laser spot, as can be seen in Fig.~\ref{fig:Laser}(b). (b,c) Corresponding weighted motion along the (b) horizontal and (c) vertical axis.}
    \label{fig:Centroid}
\end{figure}

The pointing stability of the focal spot is illustrated in Fig.~\ref{fig:Centroid}, which displays the transverse motion over time of the beam centroid in the focal plane (at the center of the target chamber). The points were taken every second, using the 10~Hz alignment beam. From it, we can conclude that the beam centroid moves at most by one spot radius.

\section{\label{sec:results}RESULTS}

\subsection{\label{sec:setup}Setup of experiment and diagnostics}

\begin{figure}[htp]
    \centering
    \includegraphics[width=0.45\textwidth]{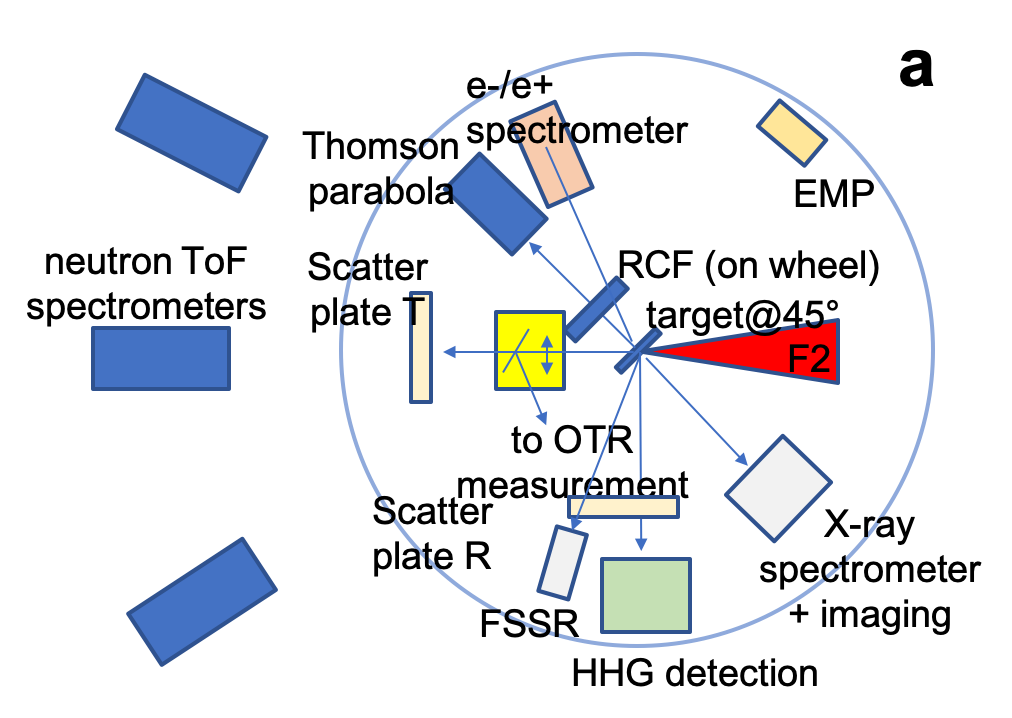}
    \includegraphics[width=0.45\textwidth]{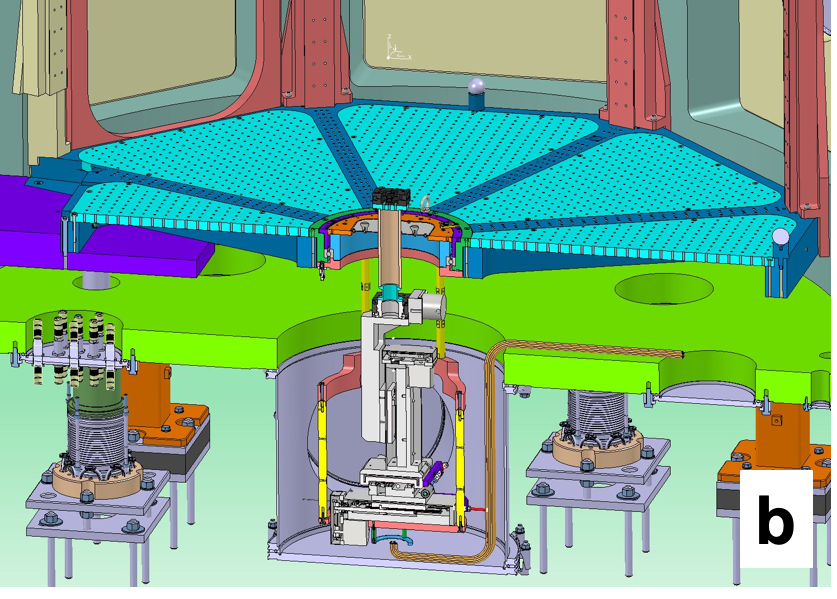}
    \caption{(a) Top-view schematic of the experimental setup. The red cone represents the focusing cone of the F2 beamline. (b) 3D rendering of the chamber bottom and target positioning system. 
    }
    \label{fig:SFA_setup}
\end{figure}

The setup of the commissioning experiment in the SFA area is sketched in Fig.~\ref{fig:SFA_setup}(a), with a view of the structure of the breadboard and target manipulator inside the chamber shown in Fig.~\ref{fig:SFA_setup}(b). As can be seen in Fig.~\ref{fig:SFA_setup}(b), the breadboard is composed of a structure (dark blue) onto which a series of triangular breadboards (turquoise) are inserted. The ensemble is decoupled from the chamber wall, ensuring that whatever is aligned in air moves minimally when the chamber is pumped down. Triangular breadboards allow users to build the setup offline and then directly insert it in the chamber when needed. The same stable structure also exists (not shown) above the equatorial plane, allowing to position diagnostics from above the target. The height of the interaction point is 40 cm above the bottom breadboard, and there are 84.5 cm between the interaction point and the roof breadboard. The target manipulator is composed of three linear stages ($xyz$), plus a rotation stage allowing for rotation around the vertical axis.

The targets were held in a rectangular holder having a regular grid structure with holes (of 3 mm diameter). The holder was composed of two parts, the targets themselves (thin foils of various metals or plastic) being held as in a sandwich between the two parts, ensuring for their planarity. The angle of laser incidence on the targets was 45°, as shown in Figure \ref{fig:SFA_setup}(a). The laser polarisation on target was tilted by $\sim$ 14° compared to the horizontal, i.e. the polarisation was close to p-polarisation. 

To characterize the laser-target interactions the following diagnostics were used downstream from the target rear: a set of RCF \cite{bolton2014} stacks on a wheel (which was motorized in and out from the target rear) to diagnose the emitted protons; a Thomson parabola (TP), using CMOS active detectors (acquired remotely) to diagnose the emitted ions; an electron spectrometer, also using CMOS active detectors and coupled with YAG:Ce stintillating crystals in order to amplify the electron signal (also acquired remotely); a scatter plate at the edge of the TP, allowing to monitor the transmitted laser energy; an off-axis parabola collecting light, in the axis of the incoming main laser beam, to send it to an optical table outside the chamber to monitor the OTR generated light by the electrons streaming from the target into vacuum \cite{doi:10.1063/1.1927328} (it is also motorized in and out from the target rear); and a series of neutron time-of-flight base modules to monitor the angular and spectral distribution of the neutrons. The CMOS detectors are RadEyes sensors, having 1" by 2" detection area, procured from Teledyne Dalsa. 

Looking at the target, the following diagnostics were used: the laser specular reflection from the target (i.e. at 90° from the incident laser) was diagnosed on a Lambertian scatter plate, which was imaged from outside the target chamber. It allowed to monitor the reflected laser energy. In the same direction, alternatively to the scatter plate, was positioned a high harmonics generation (HHG) spectrometer, allowing to register the X-ray harmonics of the reflected radiation. At angles of 10° and 58° from the target normal, were positioned two X-ray spectrometers, allowing to record plasma recombination lines. At an angle of 10° from the target normal, was also positioned an X-ray imaging system based on Fresnel zone plates.

Complementarily, we deployed a set of probes, positioned at various angles around the chamber, inside as well as outside, to measure the electromagnetic pulses (EMP) generated during the interaction.

The focused laser focal spot (either low-energy from the front-end, or amplified but attenuated before compression, see Fig. \ref{fig:Laser}(b)) was monitored by means of a motorised microscope. The X-ray imaging detailed below provides a complementary full energy spatial distribution of the laser heating on target.

\subsection{\label{sec:target}Target alignment}

\begin{figure}[htp]
    \centering
    \includegraphics[width=0.45\textwidth]{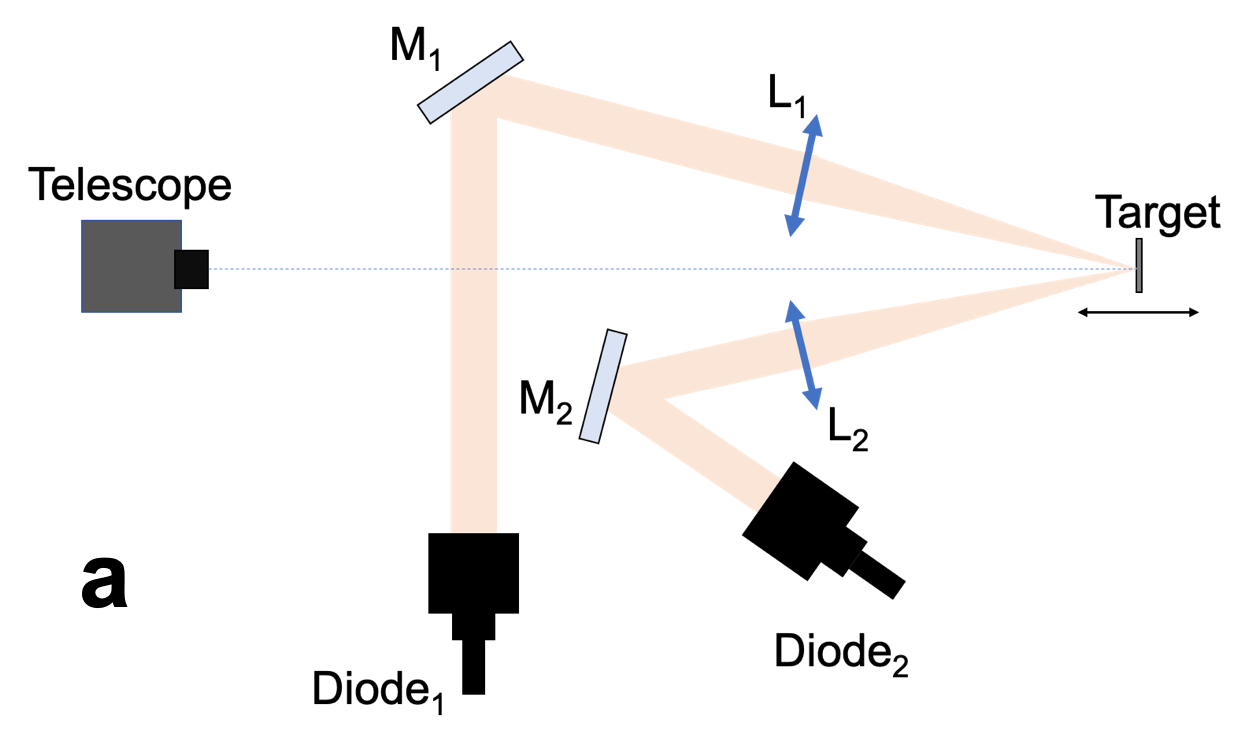}
    \includegraphics[width=0.45\textwidth]{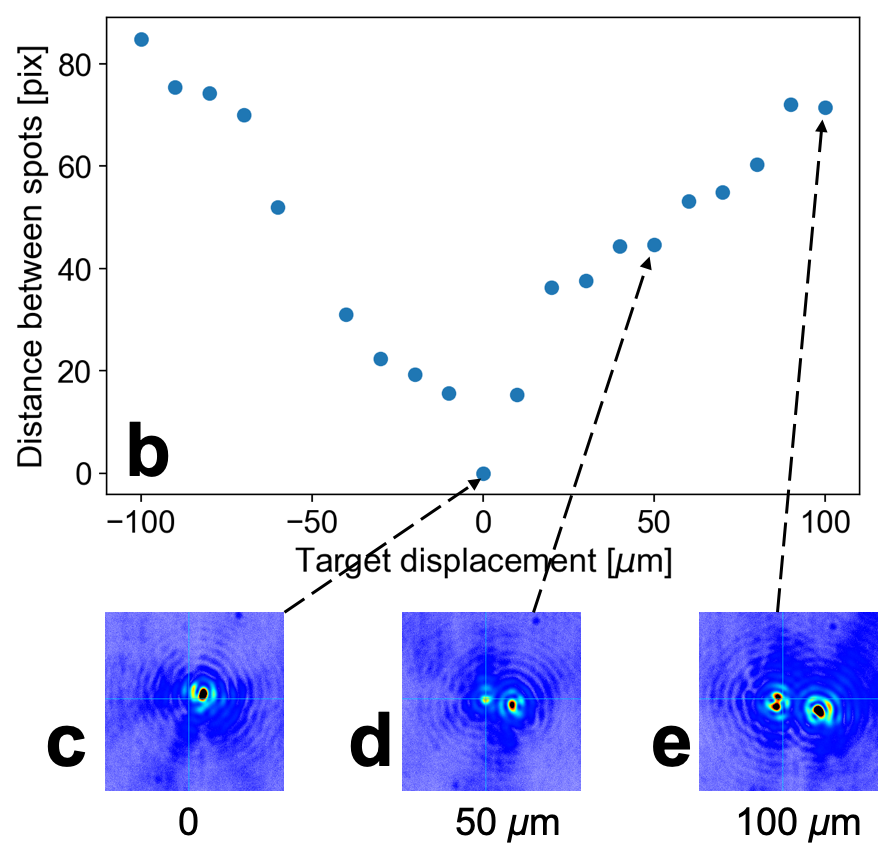}
    \caption{(a) Target alignment procedure: the alignment beams produced by laser diodes coupled to beam expanders (Diode$_1$ and Diode$_2$) are transported inside the vacuum chamber by 100\% reflective mirrors (M$_1$ and M$_2$) and are focused on the same point in space (the target chamber center, or TCC, which is position 0 in panel (b)) by means of lenses (L$_1$ and L$_2$). The target can be adjusted in focus, i.e. back and forth along the main laser axis. The telescope which is positioned outside the chamber looks at the surface of the target with a high magnification. By scanning the target along the focus axis, we can observe that the two spots of the alignment beams are either overlapped (as in panel (c)), or separated (as in panels (d) and (e)). The distance between the two spots depends on the target displacement from TCC, as shown in panel (b), where a positive direction corresponds to the target moving towards the laser focusing OAP.}
    \label{fig:SFA_alignment}
\end{figure}

The procedure of target alignment at the target chamber center (TCC) is detailed in Figure \ref{fig:SFA_alignment}. Positioning of the target in the focal plane of the laser was performed using two converging CW laser beams, delivered by two laser diodes coupled with beam expanders. Prior to positioning a target at TCC, each alignment beam was focused by a lens to the TCC point, which was materialized by the top of a wire, which was itself observed by two telescopes (from Questar-corp) positioned in air on the chamber walls, one looking down on the target over the main laser axis, the other located at 90° from the first one.

When the target surface was placed at the TCC, the telescope looking at the target above the main laser axis registered one tiny intense bright spot, as shown in Fig.~\ref{fig:SFA_alignment}(c). Moving the target back and forth led to divergence of the beam spots from the two alignment lasers, with a reduction of their intensity and increasing of their size, as shown Figs.~\ref{fig:SFA_alignment}(d) and (e). As can be seen in Fig. \ref{fig:SFA_alignment}(b), the procedure allowed each target to be easily positioned with a $\pm 20\,\rm \mu m$ precision, that is, within the Rayleigh length of the main laser beam.

\subsection{\label{sec:scatter}Scattered laser light from the target}

The scattered light in the direction of laser propagation and in the direction of specular reflection from the target surface was measured using Spectralon scatter plates and Basler CMOS cameras positioned in air on the chamber wall [see Fig.~\ref{fig:SFA_setup}(a)].

\begin{figure}[htp]
    \centering
    \includegraphics[width=0.45\textwidth]{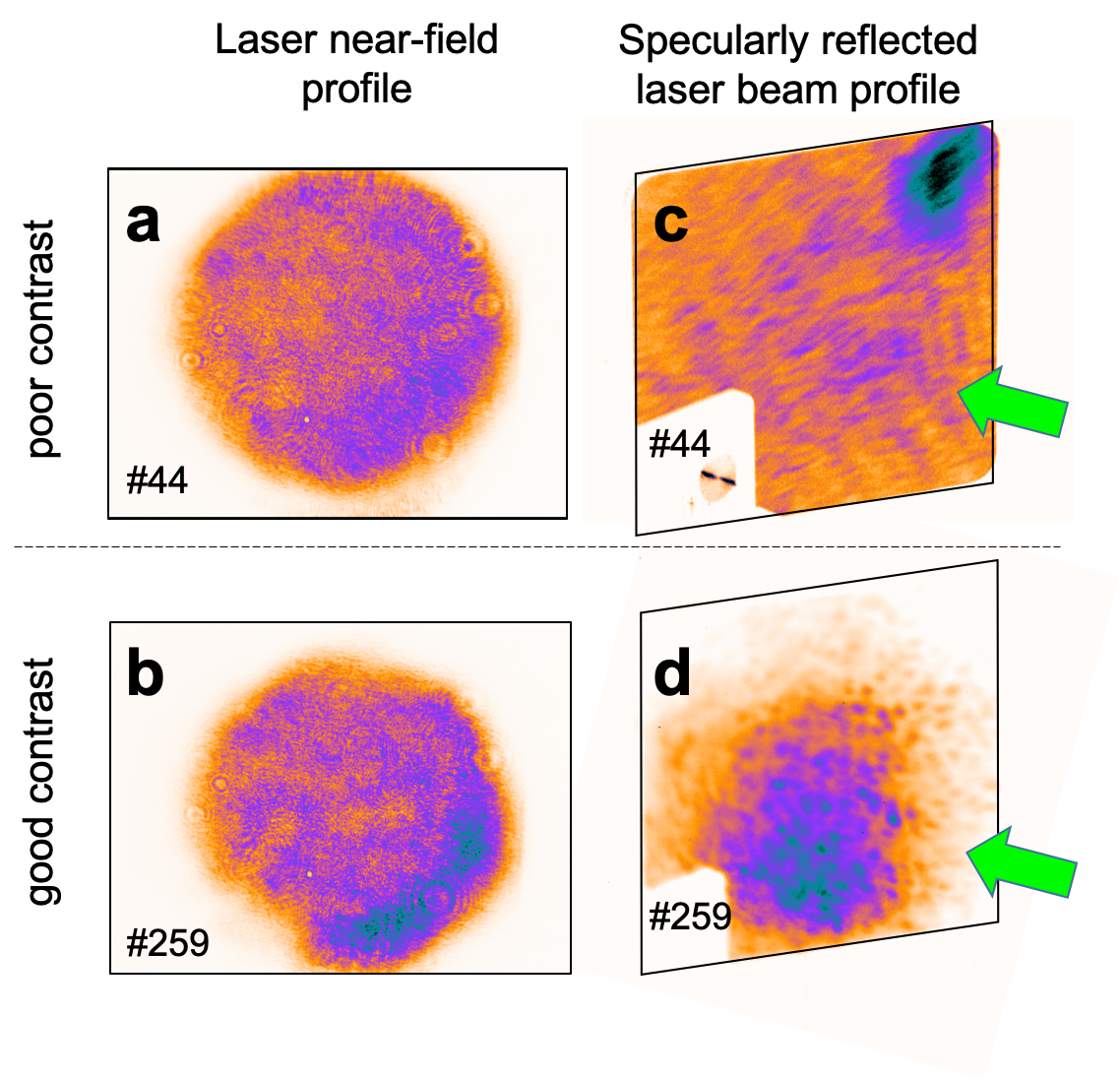}
    \caption{(a,b) Laser near-field images, captured after full amplification and before compression, and (c,d) images (as captured by an imaging camera position outside of the chamber) of the specularly reflected laser beam landing on the spectralon plate after having irradiated Al (c) 10 and (d) 3 $\mu$m thick targets. The green arrow indicates the direction of laser propagation. The white rectangle in the bottom left of the images is the shadow of an object positioned between the target and the scatter plate, partially blocking the light on way to the scatter plate. Top row: poor contrast conditions, bottom row: good contrast conditions. }
    \label{fig:Spectralon}
\end{figure}

Figure \ref{fig:Spectralon} shows the spatial distributions of the second harmonic ($2\omega = 408$ nm) of the scattered light (which was selected using a bandwidth color filter), reflected from the target. It is clearly seen, that with a good contrast the shape of the reflected beam has a similar topology as the laser near field, demonstrating a quasi mirror-like plasma surface, whereas with a poor contrast the laser beam becomes divergent and becomes widely scattered after having irradiated the target.

\subsection{\label{sec:xray}Self x-ray emission}

Three diagnostics, all looking at the target front, were implemented to characterize the laser-solid interactions in the soft X-ray domain.

\begin{figure}[htp]
    \centering
    \includegraphics[width=0.45\textwidth]{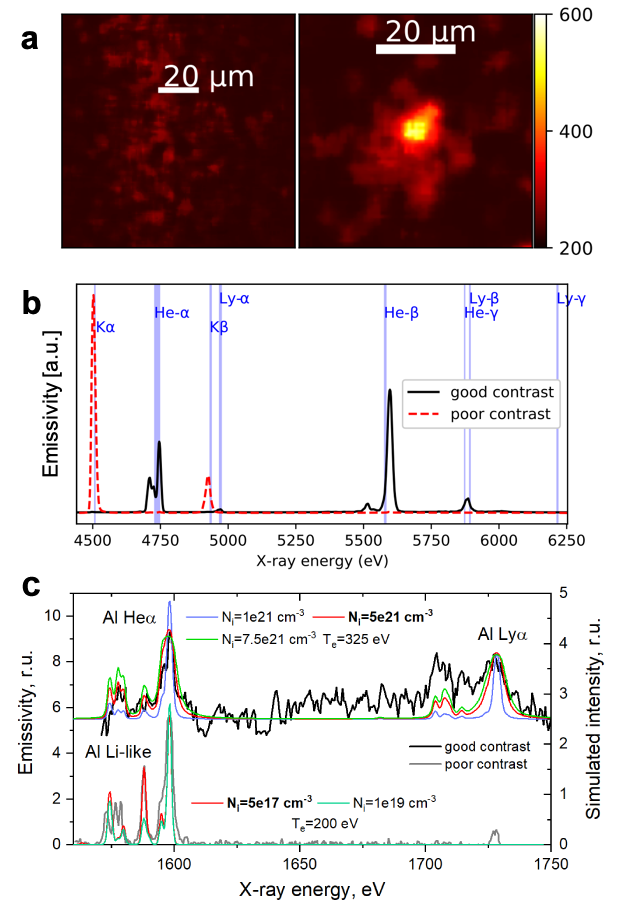}
    \caption{Self-emission X-ray signals from the target front surface. (a) Left: image from a Ti target, with poor laser contrast; (a) right: image from an Al target, with good laser contrast; (b) spectra from a thin Ti layer buried between plastic layers; (c) Al spectra obtained using the FSSR spectrometer, together with simulations performed using the PrismSPECT code. The simulated spectra are normalized to the experimental ones using the intensity of the Ly$_{\alpha}$ and He$_{\alpha}$ lines. The offset between the spectra obtained in good and poor contrast conditions is voluntary, in order to be able to distinguish one from the other.}
    \label{fig:Xray}
\end{figure}

First, the FUHRI high-resolution imager \cite{doi:10.1063/1.5042069} recorded the target self-emission at an angle of $10^\circ$ from target normal in the X-ray range $4.7 \pm 0.1\,\rm keV$. This diagnostic consists of a Fresnel zone plate of focal length $\sim 250\,\rm mm$ set with a magnification of 15.5, a multilayer mirror acting as a monochromator, and a CCD camera as a detector. The spatial resolution was previously measured between 3 and 5 $\mu$m. Figure \ref{fig:Xray}(a) illustrates typical results from this imager, highlighting the strong difference made by poor laser contrast. The < 10 $\mu$m spot of the right-hand-side panel corresponds to continuum emission from the oscillating electrons in the laser focal spot, or from their Bremsstrahlung emission, confirming the proper focusing of the laser.

The second instrument provided X-ray spectra in the Ti K-shell range (4.4 - 6.3 keV) \cite{renaudin:hal-02456723} at an angle of $10^\circ$ from target normal. It consists of a polycapillary X-ray optics able to collimate X rays over a few meters up to a curved Bragg crystal that disperses the signal on a CCD camera. This set-up allows for an active detector far away from the target in order to reduce the impact of the EMP, while keeping a good spectral resolution of E/$\Delta E$ $\sim$ 1000. A typical spectrum, from a 200 nm Ti layer buried between two $2\,\rm \mu m$ thick layers of plastic, is provided in Figure \ref{fig:Xray}(b). It can be observed that in the case of poor contrast, fast electrons accelerated by the laser in the preplasma traverse the Ti layer inducing K$_\alpha$ emission, characteristic of a weakly ionized, cold plasma. It is indeed expected that a large preplasma prevents efficient coupling between the hot electrons and the target bulk. In the case of a good contrast, the preplasma remains limited and the spectrum shows thermal emission from the hot Ti layer (He-like and H-like ions). This indicates that the temperature reached above 1 keV in the buried layer, i.e. close to the solid density. Precise atomic calculations will be carried out to interpret these results with more accuracy.

The third diagnostic was a focusing X-ray spectrometer with spatial resolution (FSSR) \cite{Faenov_1994}. It was used to record the emission of the plasma at the front surface of the target in the soft X-ray domain in a narrow spectrum range from 1.5 to 1.8~keV. The spectrometer was installed at the direction of 70$^\circ$ to target normal and 10$^\circ$ down from the equatorial plane in order to avoid the self-absorption of X-ray emission in the plasma plume and to reduce the parasitic fluorescence of the spectrometer crystal caused by hot electron flows. Figure~\ref{fig:Xray}(c) compares X-ray spectra of Aluminum as obtained in poor and good contrast conditions (black and grey curves). H-like and He-like spectral lines are observed in this range with their satellites. X-ray spectra with good and poor contrast were simulated by the radiative-collisional code PrismSPECT, taking into account an optical thickness that is equivalent to the measured focal size ($3\,\rm \mu m$). The simulation was done to fit the intensity ratios between the He$_\alpha$ line, its Li-like satellites and the Ly$_\alpha$ line, as well as the widths of the He$_\alpha$ and Ly$_\alpha$ lines. The text in bold font describes the parameters of the best fit of the simulation with respect to the experimental spectrum. The satellites of the Ly$_\alpha$ line are supposed to be generated by radiative pumping, which is not included in the model. As we see in Figure~\ref{fig:Xray}(c), when the laser temporal contrast is good, we observe an increased mean ionization charge (see the much larger ratio between the Ly$_{\alpha}$ and He$_{\alpha}$ lines) as well as broadened lines due to higher thermal electron temperature and density, respectively \cite{Andiel2002, Alkhimova2017}. With respect to the case of poor contrast, we observe that, in the case of good contrast, the ion density is increased by 4 orders of magnitude, up to $\sim 5 \times 10^{21}$~cm$^{-3}$. The electron temperature is increased as well from 200~eV to 325~eV. The spectra were recorded by both an X-ray CCD camera Andor DV430 as well as by conventional image plates. Note that in poor contrast conditions, no clear continuum emission signal from the fast electrons could be obtained, which concurs to the idea that these did not encounter much dense matter. Conversely, with good contrast, the spectrum shows a significant continuum (here subtracted from the spectrum), thus supporting the idea that the solid target was then mainly intact at the time of the high-intensity irradiation, allowing fast electrons to penetrate the solid region and produce Bremsstrahlung.

\subsection{\label{sec:hhg}High-order harmonic generation}

The high-order harmonics spectrometer coupled with a micro-channel plate detector was placed in the direction of specular light reflection from the target surface.

\begin{figure}[htp]
    \centering
    \includegraphics[width=0.45\textwidth]{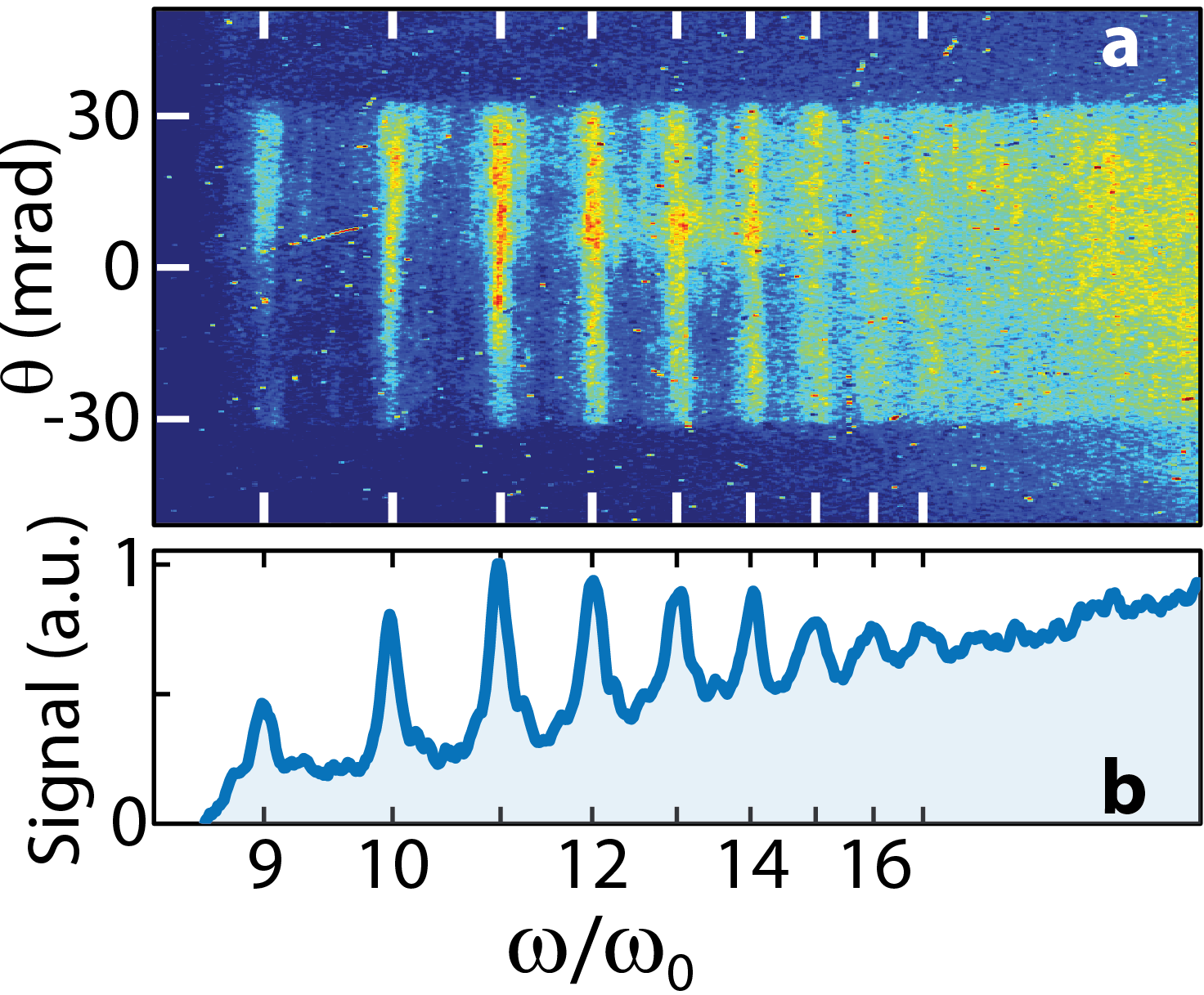}
    \caption{High-order harmonic spectrum measured during the commissioning experiment, using the F2 beamline of Apollon. XUV beam in the specular axis resolved spectrally from $\sim$ 90 to $\sim$ 10nm and resolved angularly (panel (a)) in -30 to 30 mrad. Panel (a) is the raw image measured by imaging the MCP detector with a camera. Panel (b) is the angularly integrated signal between -10 and 10 mrad.}
    \label{fig:HHG}
\end{figure}

Figure \ref{fig:HHG} presents one spectrum of high-order harmonics of the laser pulse obtained during the experiment. The obtained high harmonic spectrum extends to the harmonic 17th, i.e. to a wavelength of $\sim$ 50 nm. The laser energy in the pulse was $\sim$ 10 J, and the target was a 3 $\mu$m-thick foil of aluminum. The spectrometer is based on a XUV grating of 1200 grooves/mm from Hitachi and a micro-channel plate (MCP) detector. As illustrated in Fig.~\ref{fig:HHG}(a), the harmonic beam is angularly resolved. Note that the angular window on which the beam is observed is limited by the grating width, approximately between -30 and 30 mrad here. Due to this limitation, the harmonic beam divergence could not be evaluated.

There are two high harmonic generation (HHG) regimes when a laser is reflected off of a plasma mirror \cite{PhysRevX.9.011050, PhysRevLett.110.175001}: the coherent wake emission (CWE) and the relativistic oscillating mirror (ROM) regimes; each occurring at very different physical conditions. The CWE regime is prevalent when the plasma gradient at the target surface is shorter than $\lambda_L/20$ (where $\lambda_{L}$ is the laser wavelength), and when the laser intensity on target is not relativistic ($I < 10^{18}\,\rm Wcm^{-2}$). On the contrary, the ROM regime is prevalent when the plasma gradient scale-length is longer, between $\lambda_L/20$ and $\lambda_L/5$, and with relativistic laser intensity on target, i.e., when $I > 10^{18}\,\rm Wcm^{-2}$ such that the electrons are driven to near the speed of light \cite{salamin2006}.

The harmonic spectrum observed in Fig.~\ref{fig:HHG} most likely results from the ROM mechanism since the on-target laser intensity is evaluated to be $\sim 2 \times 10^{21}\,\rm Wcm^{-2}$, and no prior plasma mirror was used to improve the laser temporal contrast \cite{Levy:07}. Thus, a relatively extended preplasma is expected to be generated by the laser pedestal. Note that driving HHG on solid targets without using a plasma mirror is extremely challenging due the constraints imposed on the temporal contrast of the laser system, and this statement holds even at 100 TW-class facilities. The fact that we can observe HHG therefore demonstrates the excellent contrast of the laser system, consistent with other measurements.

\subsection{\label{sec:proton}Proton acceleration}

The proton beam accelerated from the target backside was characterized in energy and angle using a combination of widely accepted methods, namely a radiochromic films (RCF) stack \cite{PhysRevLett.85.2945} and a Thomson parabola spectrometer \cite{doi:10.1080/14786440208637024}. We used various target materials (plastic and metal) of thicknesses decreasing from $30\,\rm \mu m$ down to $0.8\,\rm \mu m$. Such thickness scan is a standard process in order to quantify the actual prepulse level on target. \cite{Fuchs2006natphys,PhysRevLett.99.185002,Soloviev2017}. Indeed, for relatively thick targets as explored here, target normal sheath acceleration (TNSA) is expected to be the dominant ion acceleration process \cite{doi:10.1063/1.1333697,Daido_2012,RevModPhys.85.751,PhysRevLett.116.205002}, with well-assessed characteristics (acceleration along the target normal, known scaling with laser intensity, known variation of beam divergence with energy). For this commissioning experiment, we focused only on an acceleration regime governed by TNSA, to be able to compare our results with the extensive collection of experimental data on TNSA gathered over the past two decades.

Reducing the target thickness first tends to enhance the proton energies since the sheath field set up at the backside by the widely divergent hot electrons \cite{Adam2006prl,green2008prl} is strengthened. If the target is made too thin, however, its rear surface may prematurely expand due to thermal or shock waves induced by the laser pedestal. The long density gradient then formed at the target rear is detrimental to proton acceleration \cite{Fuchs2006prl}. The target thickness maximizing proton acceleration is therefore a good indicator of the on-target prepulse level. 

We also tested double-layer targets during this campaign, namely carbon nanotube foams (CNF) deposited on Al foils. The CNF were fabricated using an improved floating catalyst chemical vapor deposition method \cite{Wang2021}. The thickness of Al foils was set to $15\,\rm \mu m$, and the thickness of the CNF ranged from $20\,\rm \mu m$ to $80\,\rm \mu m$. The density of the CNF was of $2.6 \pm 0.3\,\rm mg\,cm^{-3}$, corresponding to an electron density of $0.45 \pm 0.05n_c$ at full ioniszation ($n_c = m_e \omega_L^2\epsilon_0/e^2$ is the critical density of the plasma at the laser frequency $\omega_L=2\pi c/\lambda_L$). These yielded results (not detailed here) similar to those obtained with plain foils.

The motorized RCF stack holder was made on the base of the revolver principle, allowing to change stacks between shots remotely without access to the chamber (see Fig.~\ref{fig:SFA_photo}(b)). The stacks used RCF of type HD-v2 and EBT3 \cite{chen2016rsi}, with various filters composed of polyethylene in order to set a desirable range of energies.

\begin{figure}[htp]
    \centering
    \includegraphics[width=0.45\textwidth]{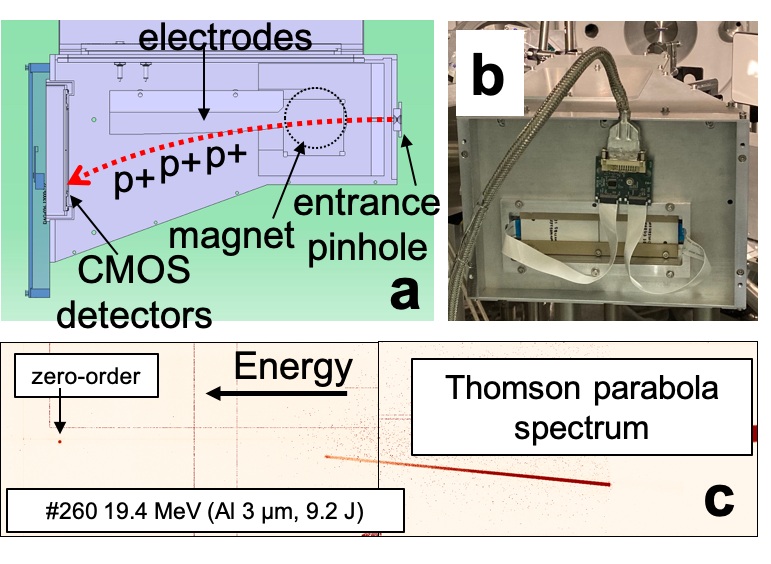}
    \caption{(a) Thomson parabola (TP) schematic top view; (b) photograph of the back view with RadEye CMOSes; (c) typical spectral image obtained in a single shot with the TP. One clearly sees the boundary between the two CMOS detectors that are mounted head to tail. The low energy part of the spectrum has been intentionally cut by a thick Al filter since it saturated the detector, as the number of particles increases exponentially toward the low energies.
    }
    \label{fig:TP}
\end{figure}

We conceived the electron/Thomson parabola spectrometer ANNA (ApolloN thomsoN parabolA) tailored to the needs of particle acceleration experiments at Apollon. Once passed through the entrance pinhole (whose diameter may span between 50 µm and 1 mm) particles are dispersed by a 1.1 T magnetic field (produced by 70 mm diameter round magnets, see Fig. \ref{fig:SFA_photo}(a)) and deflected by an electric field between two copper slabs whose voltage difference can rise up to 10 kV. An opportune combination of pinhole diameter, voltage difference and geometrical shape of the slabs allows to record the entire proton spectral distribution between 1 and 100 MeV. Three different backs have been designed in order to collect (protons and) ions on a film detector (Image Plate or CR-39), a MCP or a CMOS detector. Moreover, electron spectral distribution between 4 and 100 MeV can optionally be obtained by collecting the fluorescence of calibrated YAG(Ce) crystals  placed on one side of ANNA (normal to the ions detection plan) and imaged by a camera.
ANNA was positioned along the target normal, with its front pinhole at a distance of 320 mm from TCC, and with a pinhole of 200 µm diameter. The electrodes were charged at $\pm 2$ kV. To operate in high-repetition mode, the TP detector was composed of two highly sensitive RadEye CMOS matrices that were mounted head to tail. Protons impacted the CMOS surface directly. Figure~\ref{fig:TP}(a) presents the schematic top-view of the spectrometer and Figure~\ref{fig:TP}(b) shows the mounting and cabling of the CMOS detectors inside the TP assembly; an additional cover was added on top of what is shown in the photo in order to have the detectors enclosed in a Faraday cage, to avoid electromagnetic perturbations onto the readout system. A detailed calibration of the instrument and of the CMOS detectors was performed, but since it is out of scope of this report, it will be reported elsewhere. 

The CMOS detectors were triggered 1 ms before the shot and readout right after. The signal was acquired remotely after each shot allowing to use the TP in multi-shot regime, thus removing the need to have an access to the detector, as is the case when employing commonly used imaging plates \cite{Lelasseux2020}, but which imposes to  break the vacuum of the target chamber.

Figure~\ref{fig:TP}(c) shows a typical proton spectrum, with a 19.4 MeV cutoff energy, emitted from a 3 $\mu$m aluminum target irradiated with an on-target pulse energy of 9.2 J, as registered by the Thomson parabola spectrometer. The measurement consists of two separated images collected by two RadEye CMOS matrixes. The so-called zero-order is the point-projection of the neutral atoms and x-rays emitted by the target onto the detector. It serves as a reference from which the deflection of the charged particles can be accounted for. A track of deflected protons can be seen on the right-hand side of the image, with a  high-energy cut-off and a broad spectrum, which are both typical signatures of TNSA acceleration \cite{Daido_2012}. The proton track is parabolically curved, as is expected \cite{doi:10.1080/14786440208637024}.

We have been able to generate proton beams with maximum cutoff energy around 28 MeV from 2-3 $\mu$m thick Al foils, consistent with results obtained at other facilities under similar experimental conditions \cite{Ogura:12,Soloviev2017}.  
%Lower thicknesses are damaged by the ASE and the ramp-up of intensity leading to the 25 fs main peak. These can be however easily suppressed by the use of a plasma mirror for the future campaigns.

\begin{figure}[htp]
    \centering
    \includegraphics[width=0.45\textwidth]{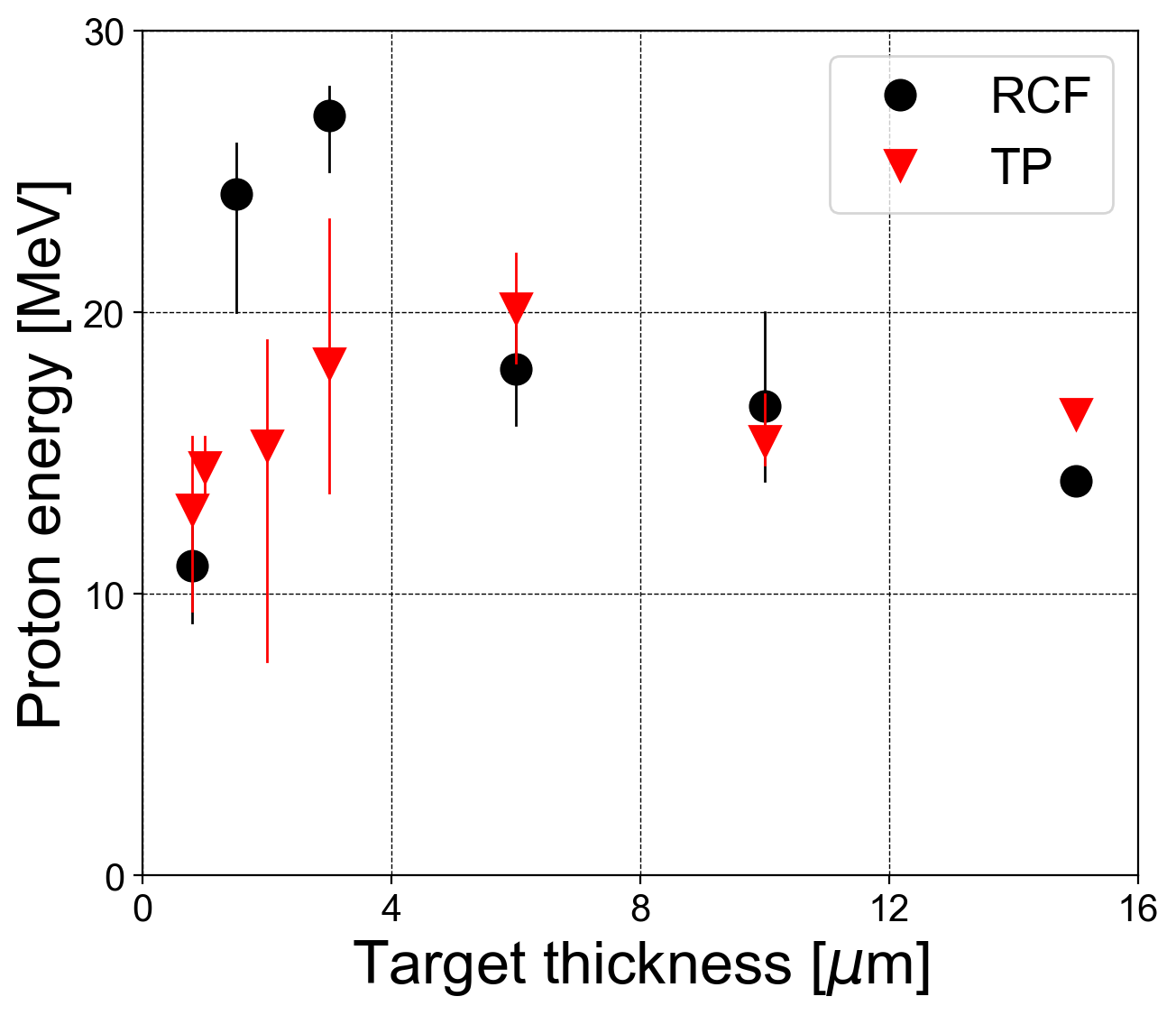}
    \caption{Maximum proton energies, as registered by RCFs and Thomson parabola, depending on the target thickness. The targets are plain Al foils.
    }
    \label{fig:Protons}
\end{figure}

Figure \ref{fig:Protons} plots variations in the proton cutoff energy with the Al foil thickness, as recorded by the RCF and TP diagnostics. Black dots for RCF and red triangles for TP represent the averaged proton cutoff values over several shots and the vertical lines presents the range of the recorded cut-off energy values over these shots. 

The proton cutoff energies recorded by the RCF appear to be higher than those diagnosed by the TP for 1.5 and $3\,\rm \mu m$ target thicknesses. This is due to deflection of the proton beam, as observed on the RCF. The higher energy portion of the proton beam turned out to deviate from the target normal direction and therefore was not collected by the TP. This phenomenon can result from a hybrid acceleration mechanism, as proposed in Ref.~\cite{wagner2016prl}, yet under quite distinct interaction conditions. A more thorough analysis is thus needed to interpret our data, which will be the subject of a separate paper. 

\subsection{\label{sec:electron}Hot-electron generation}

The electron spectrometer was placed with its entrance pinhole at 440 mm from TCC, along the axis perpendicular to the target surface, alternatively to the TP. The entrance pinhole diameter was 1 mm. The spectrometer held a pair of magnets that created a magnetic field of 1 T to disperse the incoming electrons, in space, with energy from 5 MeV to 100 MeV.

\begin{figure}[htp]
    \centering
    \includegraphics[width=0.45\textwidth]{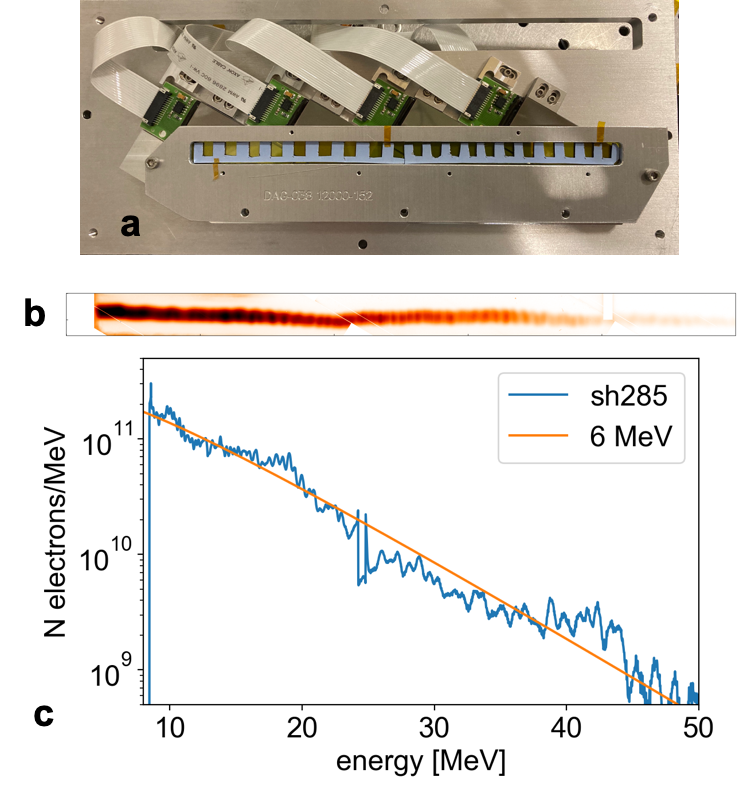}
    \caption{(a) Photograph of the lateral panel of the electron spectrometer detector assembly, showing how the RadEye CMOSes detectors are mounted on it; (b) raw image of a recorded electron spectrum from composite 20 nm aluminum $+$ 23 $\mu$m Polyethylene terephthalate (PET) target. For this particular shot the distance from TCC to spectrometer entrance was 290 mm. (c) Hot electron spectrum retrieved from the raw image (shown in (b)) and the corresponding Maxwellian distribution fitting using an energy of 6 MeV.  
    }
    \label{fig:Electrons}
\end{figure}

The detector of the spectrometer was composed of four RadEye CMOS detectors, as shown in Figure \ref{fig:Electrons} (a), which were installed at an angle to maximize the use of the detector and to accommodate the fixed position of the electronic board. A 1 mm thick YAG:Ce scintillating crystal was set flush onto the detectors to convert the incoming electron signal into visible-light at central wavelength of 550 nm (see Figure \ref{fig:Electrons} (a)). Like the TP diagnostic, the electron spectrometer was used in a multi-shot operation mode, i.e. the RadEye signal was read remotely after every shot.

To have an absolute calibration of the electron deposition on the CMOS, for one shot we placed a grid made of an image plate with a known response to the electron impact \cite{chen2008rsi} in front the YAG:Ce scintillator, as can be seen in Figure \ref{fig:Electrons} (a). Detailed review of the absolute calibration of the RadEye CMOS coupled with a Yag:Ce crystal will be the subject of a separated technical publication.

The raw signal registered by the RadEye is presented in Figure \ref{fig:Electrons} (b), and the extracted electron spectra is shown in Figure \ref{fig:Electrons} (c).

\subsection{\label{sec:neutron}Neutron generation}

The spectra of neutrons produced by converting the accelerated protons into a converter (Li or Pb) \cite{Alejo2015} was diagnosed with an array of time-of-flight modules positioned outside of the target chamber (see Figure \ref{fig:SFA_setup} (a)). These modules are EJ-254 with a 1\% loading of boron scintillator rods coupled, at both ends of the rods to two photomultiplier tubes (PMT) and ran in coincidence counting mode.

\begin{figure}[htp]
    \centering
    \includegraphics[width=0.45\textwidth]{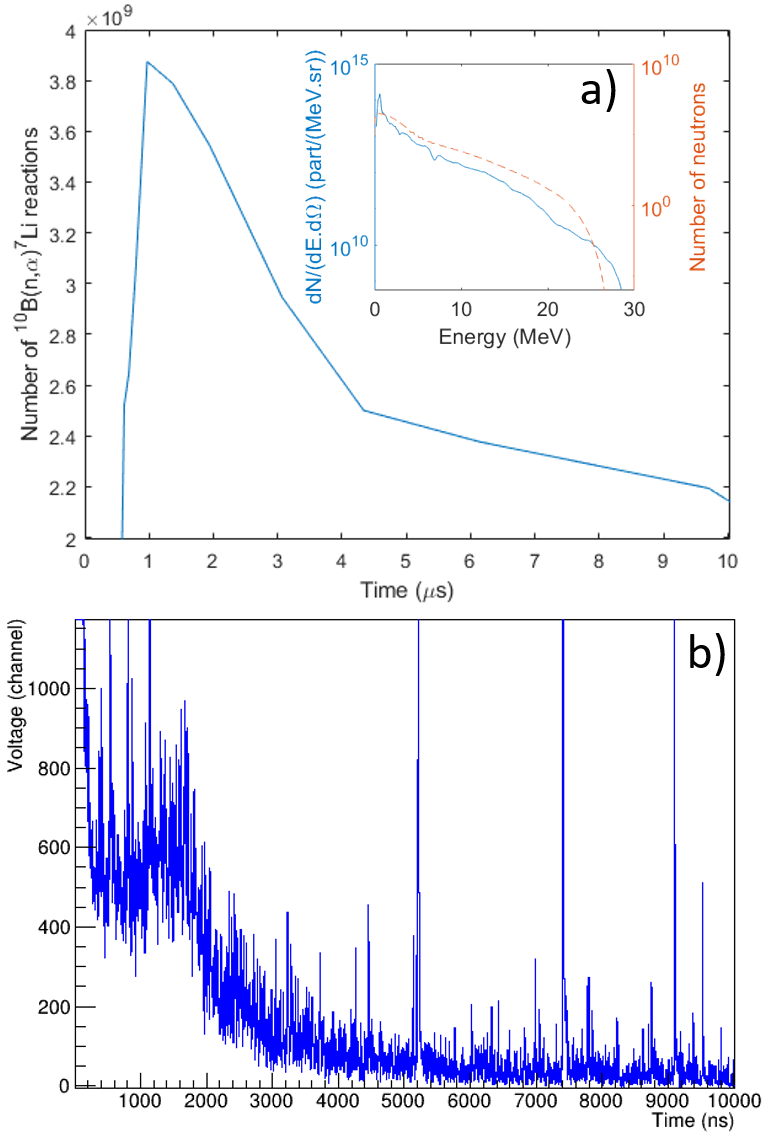}
    \caption{(a) Predicted number of events (as a function of time) produced in the boron-doped plastic scintillator of type EJ-254, placed at 6 m from the target, when exposed to the neutron spectrum shown in dashed line in the inset. That neutron spectrum results from having a proton beam with maximum energy 28 MeV, also shown in the inset, in full line, irradiating a Pb converter. The neutron generation in Pb is simulated using GEANT4. The events in the scintillator are simulated using the known cross-section in 10B (which decreases strongly when the incoming neutron energy increases) and the neutron spectrum (dashed line) is shown in the inset: starting at t=0, there are no events, because there are no neutrons that have flown yet to the detector. As the neutron spectrum peaks around 1 MeV, and decreases for lower energies, this results in the events having a peak around 1 $\mu$s. (b) Actual time-of-flight chronogram recorded during the commissioning experiment, from a proton beam with maximum energy 28 MeV impinging on an 1 mm thick Pb converter. The signal is high at t=0 due to the gamma flash and the EMP. However, we notice a clear “bump” located at $\sim$ 1.5 $\mu$s, akin to what is expected based on the calculation shown in panel (a).}
    \label{fig:nTOF}
\end{figure}

These have been shown to perform well in high-power laser electromagnetic environments \cite{Mirfayzi2015rsi}. We looked at the target at various angles. An example of such measurement is shown in Figure \ref{fig:nTOF}. A detailed analysis of the signals is undergoing and will be reported separately.

\subsection{\label{sec:otr}Optical transition radiation}

Diagnostics of the coherent optical transition radiation (OTR), emitted by the hot electrons when they cross the rear side of the solid target toward vacuum \cite{doi:10.1063/1.1927328}, was also implemented in the experimental setup. A particular interest of this diagnostic is that it can be used to image the electron generation at full power, and thus constitutes essentially a monitor of real intensity distribution of the high-power laser focal spot.

The OTR light was collected by means of an of-axis parabolic mirror ($f/6$), placed on the same axis as the main laser beam and 150 mm downstream from the target (see Fig. \ref{fig:SFA_setup}). The light originating from the target was collimated by the parabola and transported from the vacuum chamber to an optical table, where it was analyzed by a set of Andor cameras. These registered the spatial distribution of the radiation at the fundamental (1$\omega$ = 815 nm) and second harmonic (2$\omega$ = 408 nm) wavelengths of the laser. The cameras imaged the plane of the target positioned at TCC.
% Narrow-band filters were used for harmonics isolation. 
Additionally, another Andor camera coupled with a spectrometer was also used, allowing to register the spectral components of CTR harmonics.
To prevent the collection OAP mirror from damages it was protected by a neutral optical density filter.

\begin{figure}[htp]
    \centering
    \includegraphics[width=0.30\textwidth]{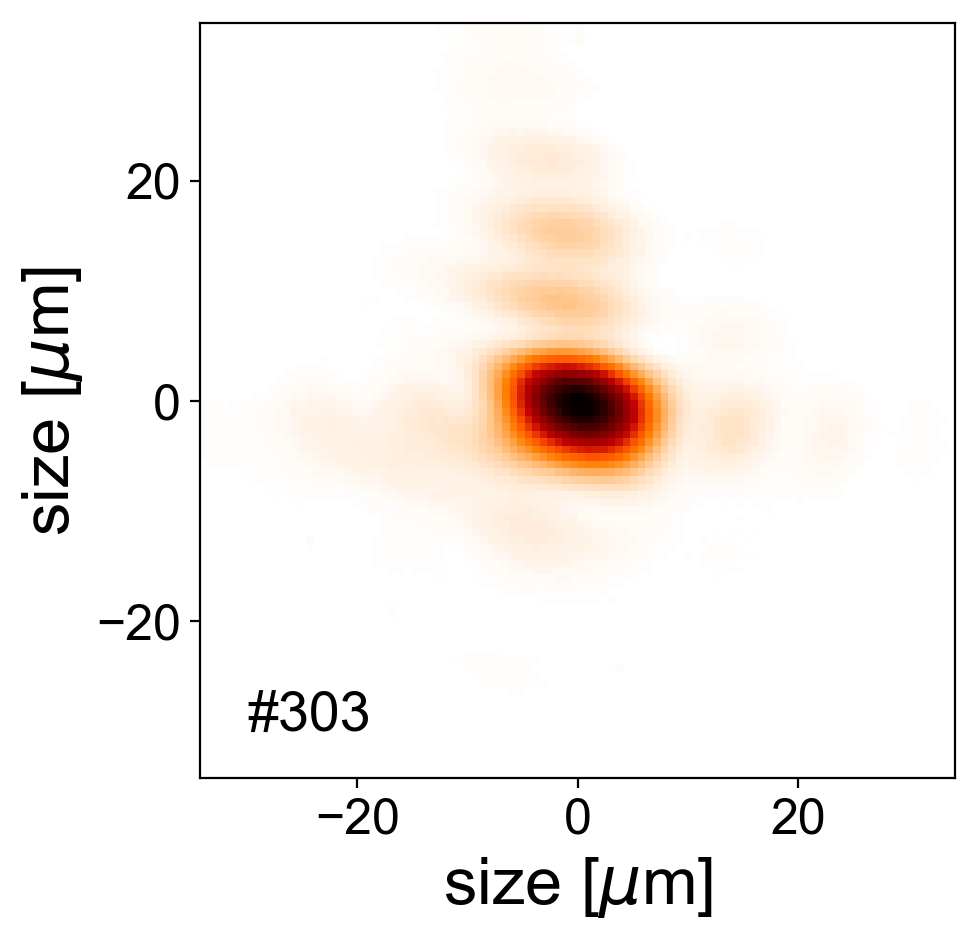}
    \caption{OTR image collected at 1$\omega$ and emitted from the rear side of a 2 $\mu$m thick Al target. The image is larger than that of the focal spot shown in Fig. \ref{fig:Laser} (b) due to the intrinsic limited resolution of the f/6 imaging parabola.}
    \label{fig:otr}
\end{figure}

Figure \ref{fig:otr} demonstrates the raw spatial distribution of the OTR as recorded at the fundamental laser harmonic, registered while irradiating a 2 $\mu$m thick aluminum target. The beam-spot shape is well reproducible from shot to shot. It basically shows that the zone from which the hot electrons are generated is narrow and well-localized. Forthcoming detailed analysis of the OTR results will be a subject of a separated work.

\subsection{\label{sec:emp}Electromagnetic pulse generation}

Monitoring the levels of  Electromagnetic Pulses (EMPs) is of paramount importance for the development of high-power, high-energy laser facilities \cite{Dubois_PRE, Consoli_HPLSE, Consoli_Phtransa}. These transient electromagnetic fields in the radiofrequency-microwave range (MHz to tens of GHz), can reach remarkable intensities (beyond the MV/m level close to TCC and up to $\sim$1 m away from it) that scale with laser energy and mostly with laser intensity at focus \cite{Consoli_HPLSE}, and for this reason can be a source of fatal damage or failure of electrical equipment. On the other hand, they may represent an interesting diagnostic tool of laser plasma interaction since the intensity, the temporal and the spectral features of the related electric and magnetic fields can be correlated to the specific interaction conditions of each shot series of the experimental campaign. Detecting the laser-generated EMP radiation is then of primary importance.

 To this purpose, the Apollon facility has been equipped with an EMP measurement platform consisting of the following elements: 
\begin{itemize}
    \item[i)] two {\it D-dot} probes Prodyn AD-80D(R), differential electric field sensors \cite{DDOT,Consoli_HPLSE};
    \item[ii)] two {\it B-dot} probes Prodyn RB-230, differential magnetic field sensors \cite{BDOT,Consoli_HPLSE};
    \item[iii)] a {\it balun} Prodyn B170 associated to each probe, for differential signal detection and rejection of common-mode signals \cite{BALUN,Consoli_HPLSE};
    \item[iv)] a suitable set of semi-rigid, double shielded cables; 
    \item[v)] a high performance Faraday cage: Siepel (French supplier) with attenuation > 100 dB from 30 MHz to 6 GHz; 
    \item[vi)] an oscilloscope Agilent  Infinium 90804A (4 channels, 8 GHz).
\end{itemize}

\begin{figure}[htp]
    \centering
    \includegraphics[width=0.45\textwidth]{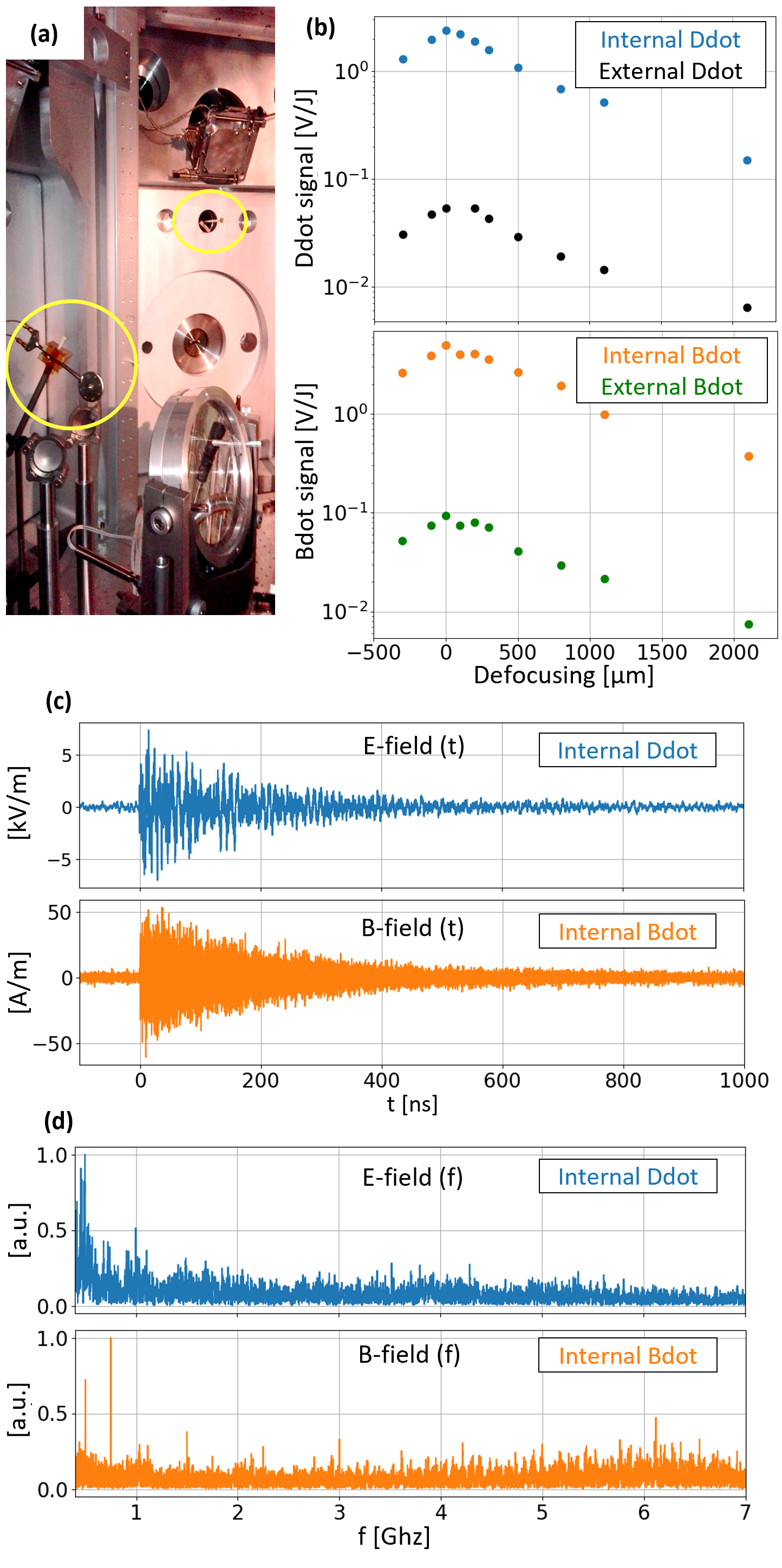}
    \caption{(a) Photograph of the electromagnetic field probes inside the target chamber: the D-dot is on the left behind the focusing parabola, and the B-dot is on the right. (b) Amplitude of the raw signals (expressed in Volt, and normalized to the laser energy), as a function of the laser focusing position on target (a motion in the positive direction means that the focusing parabola was moved closer to the target). (c) Temporal evolution of the electric (blue, D-dot, V/m) and magnetic (orange, B-dot, A/m) fields inside the target chamber. (d) Frequency spectra of the electric and magnetic fields displayed in (c).}
    \label{fig:emp}
\end{figure}

In this commissioning campaign, one of the D-dot probes was placed inside the chamber at a distance of $\sim$65 cm from TCC and behind the focusing parabola, which acted as a suitable shield for direct particles and ionizing electromagnetic radiation coming from the plasma, as shown in Fig.\ref{fig:emp}(a). A second D-dot probe was placed outside the chamber at a distance of $\sim$370 cm from  TCC. 
Similarly, a B-dot sensor was placed inside the chamber (see Fig.\ref{fig:emp}(a)) at a distance of $\sim$100 cm from TCC and shielded by a black Al foil ($\sim$100 $\mu$m), and a second outside the chamber at $\sim$320 cm from TCC.

Both of these probes (D-dot and B-dot) give information on the time derivative of the detected electric and magnetic fields. An appropriate numerical integration is thus required to retrieve information on the fields from the stored signals. 
Characteristic signals, for both the electric and the magnetic fields, obtained in this experimental campaign by the probes inside the chamber, are reported in Fig.\ref{fig:emp}(c) and \ref{fig:emp}(d), in the time and frequency domains, respectively. Typical maximum intensities for the measured electric fields when shooting on aluminum targets of 2 $\mu$m thickness were: $\sim$10 kV/m peak-to-peak inside the vacuum chamber (see Fig.\ref{fig:emp}(c)), and $\sim100$ V/m peak-to-peak outside. Typical H field values were: 60 A/m inside (see Fig.\ref{fig:emp}(c)) and 1.5 A/m outside the chamber. The EMP duration was of the order of several hundreds of nanoseconds. EMPs are inversely dependent on the distance from TCC; thus at positions closer to the interaction point, fields much higher than those detected by the internal D-dot are produced. These exceed the typical damage threshold (10 kV/m) for unshielded active electronic equipment. 

An intensity scan, performed by changing the focusing position of the parabola at fairly constant laser pulse energy, is shown in Fig.\ref{fig:emp}(b). Here the maximum values of the signals on the oscilloscope are expressed in Volt and normalized to the laser energy. As expected, the measured maximum intensity of both electric and magnetic fields sensed both inside and outside the chamber was obtained at the best focus, corresponding to maximum laser intensity.

These tests, conducted at laser intensities that are still lower than the future full-power output of Apollon, have been fundamental to introduce an EMP measurement  platform and start optimizing shielding and mitigation techniques and management in signal transmission. In fact, already at these laser intensities, EMPs are recognized to be serious threats for electronics inside the chamber. 
The Apollon upgrade to the multi-PW level will allow to enter a regime of EMP emissions still unexplored, and supply data so far lacking for short and energetic pulses.
The general scaling reported from experimental data of other existing laser facilities \cite{Consoli_HPLSE}, shows that for 100 J laser pulses in the picosecond regime EMP fields grow up to several hundreds of kV/m at $\sim 1$ m from TCC. 
Thus, when Apollon will run in the multi-PW regime, we expect EMP levels to increase by more than one order of magnitude with respect to what was detected in this commissioning campaign. However, since EMP power depends also on the specific laser intensity at focus and target used, it is necessary to be well prepared to even more significant growths of these pulses.

\section{Conclusions and perspectives} \label{sec:conclusion}

The Apollon laser facility is a users' facility open to researchers worldwide. The present report details how its SFA area underwent successful commissioning. A number of diagnostic techniques operating in a multi-shot regime were tested. We demonstrated that TNSA proton beams with cut-off energies energy around 28 MeV could be produced from 2-3 $\mu$m thick Al foils, which are well suitable for applications, e.g. proton radiography. The emissions of electrons, ions and high energy electromagnetic radiation that were recorded show good laser-target coupling and an overall performance that is very consistent with what has been reported by similar international facilities. We show that the laser displays very good temporal contrast characteristics, and that the plasma gradient scale-length at the target front is between $\lambda_{L}$/20 and $\lambda_{L}$/5. Upcoming experimental campaigns will be performed to test the implementation of plasma mirrors \cite{Thaury2007natphys} in order to allow interactions in a ultra-high contrast regime, or with ultra-thin targets. The next phase will be the upgrade of the Apollon laser to 4 PW, which is scheduled for the year 2022, and which will be followed by a further upgrade to the final 10 PW level.

\section{\label{sec:acknow}ACKNOWLEDGMENTS}
The authors would like to thank all teams of the laboratories which contributed to the success of the facility, that is all the CILEX consortium, which was established to build Apollon. This work was supported by funding from the European Research Council (ERC) under the European Unions Horizon 2020 research and innovation program (Grant Agreement No. 787539, project GENESIS), and by Grant ANR-17-CE30- 0026-Pinnacle from Agence Nationale de la Recherche. We acknowledge, in the frame of the GENESIS project, the support provided by Extreme Light Infrastructure Nuclear Physics (ELI- NP) Phase II, a project co-financed by the Romanian Government and European Union through the European Regional Development Fund, and by the project $\ ELI-RO-2020-23$ funded by IFA (Romania), to design, build and test the neutron detectors used in this project, as well as parts of the OTR diagnostic. JIHT RAS team members are supported by the Ministry of Science and Higher Education of the Russian Federation (State Assignment No. 075-00460-21-00). The reported study was also funded by the Russian Foundation for Basic Research, projects No. 20-02-00790.

\bibliographystyle{apsrev4-2-titles}
\bibliography{Apollon}% Produces the bibliography via BibTeX.

%apsrev4-2.bst 2019-01-14 (MD) hand-edited version of apsrev4-1.bst
%Control: key (0)
%Control: author (72) initials jnrlst
%Control: editor formatted (1) identically to author
%Control: production of article title (-1) disabled
%Control: page (1) range
%Control: year (1) truncated
%Control: production of eprint (0) enabled
\begin{thebibliography}{46}%
\makeatletter
\providecommand \@ifxundefined [1]{%
 \@ifx{#1\undefined}
}%
\providecommand \@ifnum [1]{%
 \ifnum #1\expandafter \@firstoftwo
 \else \expandafter \@secondoftwo
 \fi
}%
\providecommand \@ifx [1]{%
 \ifx #1\expandafter \@firstoftwo
 \else \expandafter \@secondoftwo
 \fi
}%
\providecommand \natexlab [1]{#1}%
\providecommand \enquote  [1]{``#1''}%
\providecommand \bibnamefont  [1]{#1}%
\providecommand \bibfnamefont [1]{#1}%
\providecommand \citenamefont [1]{#1}%
\providecommand \href@noop [0]{\@secondoftwo}%
\providecommand \href [0]{\begingroup \@sanitize@url \@href}%
\providecommand \@href[1]{\@@startlink{#1}\@@href}%
\providecommand \@@href[1]{\endgroup#1\@@endlink}%
\providecommand \@sanitize@url [0]{\catcode `\\12\catcode `\$12\catcode
  `\&12\catcode `\#12\catcode `\^12\catcode `\_12\catcode `\%12\relax}%
\providecommand \@@startlink[1]{}%
\providecommand \@@endlink[0]{}%
\providecommand \url  [0]{\begingroup\@sanitize@url \@url }%
\providecommand \@url [1]{\endgroup\@href {#1}{\urlprefix }}%
\providecommand \urlprefix  [0]{URL }%
\providecommand \Eprint [0]{\href }%
\providecommand \doibase [0]{https://doi.org/}%
\providecommand \selectlanguage [0]{\@gobble}%
\providecommand \bibinfo  [0]{\@secondoftwo}%
\providecommand \bibfield  [0]{\@secondoftwo}%
\providecommand \translation [1]{[#1]}%
\providecommand \BibitemOpen [0]{}%
\providecommand \bibitemStop [0]{}%
\providecommand \bibitemNoStop [0]{.\EOS\space}%
\providecommand \EOS [0]{\spacefactor3000\relax}%
\providecommand \BibitemShut  [1]{\csname bibitem#1\endcsname}%
\let\auto@bib@innerbib\@empty
%</preamble>
\bibitem [{\citenamefont {Danson}\ \emph {et~al.}(2019)\citenamefont {Danson},
  \citenamefont {Haefner}, \citenamefont {Bromage}, \citenamefont {Butcher},
  \citenamefont {Chanteloup}, \citenamefont {Chowdhury}, \citenamefont
  {Galvanauskas}, \citenamefont {Gizzi}, \citenamefont {Hein}, \citenamefont
  {Hillier},\ and\ \citenamefont {et~al.}}]{danson_2019}%
  \BibitemOpen
  \bibfield  {author} {\bibinfo {author} {\bibfnamefont {C.~N.}\ \bibnamefont
  {Danson}}, \bibinfo {author} {\bibfnamefont {C.}~\bibnamefont {Haefner}},
  \bibinfo {author} {\bibfnamefont {J.}~\bibnamefont {Bromage}}, \bibinfo
  {author} {\bibfnamefont {T.}~\bibnamefont {Butcher}}, \bibinfo {author}
  {\bibfnamefont {J.-C.~F.}\ \bibnamefont {Chanteloup}}, \bibinfo {author}
  {\bibfnamefont {E.~A.}\ \bibnamefont {Chowdhury}}, \bibinfo {author}
  {\bibfnamefont {A.}~\bibnamefont {Galvanauskas}}, \bibinfo {author}
  {\bibfnamefont {L.~A.}\ \bibnamefont {Gizzi}}, \bibinfo {author}
  {\bibfnamefont {J.}~\bibnamefont {Hein}}, \bibinfo {author} {\bibfnamefont
  {D.~I.}\ \bibnamefont {Hillier}},\ and\ \bibinfo {author} {\bibnamefont
  {et~al.}},\ }\bibfield  {title} {\emph {\bibinfo {title} {Petawatt and
  exawatt class lasers worldwide}},\ }\href
  {https://doi.org/10.1017/hpl.2019.36} {\bibfield  {journal} {\bibinfo
  {journal} {High Power Laser Science and Engineering}\ }\textbf {\bibinfo
  {volume} {7}},\ \bibinfo {pages} {e54} (\bibinfo {year} {2019})}\BibitemShut
  {NoStop}%
\bibitem [{\citenamefont {Lureau}\ \emph {et~al.}(2020)\citenamefont {Lureau},
  \citenamefont {Matras}, \citenamefont {Chalus}, \citenamefont {Derycke},
  \citenamefont {Morbieu}, \citenamefont {Radier}, \citenamefont {Casagrande},
  \citenamefont {Laux}, \citenamefont {Ricaud}, \citenamefont {Rey},\ and\
  \citenamefont {et~al.}}]{lureau_2020}%
  \BibitemOpen
  \bibfield  {author} {\bibinfo {author} {\bibfnamefont {F.}~\bibnamefont
  {Lureau}}, \bibinfo {author} {\bibfnamefont {G.}~\bibnamefont {Matras}},
  \bibinfo {author} {\bibfnamefont {O.}~\bibnamefont {Chalus}}, \bibinfo
  {author} {\bibfnamefont {C.}~\bibnamefont {Derycke}}, \bibinfo {author}
  {\bibfnamefont {T.}~\bibnamefont {Morbieu}}, \bibinfo {author} {\bibfnamefont
  {C.}~\bibnamefont {Radier}}, \bibinfo {author} {\bibfnamefont
  {O.}~\bibnamefont {Casagrande}}, \bibinfo {author} {\bibfnamefont
  {S.}~\bibnamefont {Laux}}, \bibinfo {author} {\bibfnamefont {S.}~\bibnamefont
  {Ricaud}}, \bibinfo {author} {\bibfnamefont {G.}~\bibnamefont {Rey}},\ and\
  \bibinfo {author} {\bibnamefont {et~al.}},\ }\bibfield  {title} {\emph
  {\bibinfo {title} {High-energy hybrid femtosecond laser system demonstrating
  2 × 10 PW capability}},\ }\href {https://doi.org/10.1017/hpl.2020.41}
  {\bibfield  {journal} {\bibinfo  {journal} {High Power Laser Science and
  Engineering}\ }\textbf {\bibinfo {volume} {8}},\ \bibinfo {pages} {e43}
  (\bibinfo {year} {2020})}\BibitemShut {NoStop}%
\bibitem [{\citenamefont {Papadopoulos}\ \emph {et~al.}(2017)\citenamefont
  {Papadopoulos}, \citenamefont {Ramirez}, \citenamefont {Genevrier},
  \citenamefont {Ranc}, \citenamefont {Lebas}, \citenamefont {Pellegrina},
  \citenamefont {Blanc}, \citenamefont {Monot}, \citenamefont {Martin},
  \citenamefont {Zou}, \citenamefont {Mathieu}, \citenamefont {Audebert},
  \citenamefont {Georges},\ and\ \citenamefont {Druon}}]{Papadopoulos:17}%
  \BibitemOpen
  \bibfield  {author} {\bibinfo {author} {\bibfnamefont {D.~N.}\ \bibnamefont
  {Papadopoulos}}, \bibinfo {author} {\bibfnamefont {P.}~\bibnamefont
  {Ramirez}}, \bibinfo {author} {\bibfnamefont {K.}~\bibnamefont {Genevrier}},
  \bibinfo {author} {\bibfnamefont {L.}~\bibnamefont {Ranc}}, \bibinfo {author}
  {\bibfnamefont {N.}~\bibnamefont {Lebas}}, \bibinfo {author} {\bibfnamefont
  {A.}~\bibnamefont {Pellegrina}}, \bibinfo {author} {\bibfnamefont {C.~L.}\
  \bibnamefont {Blanc}}, \bibinfo {author} {\bibfnamefont {P.}~\bibnamefont
  {Monot}}, \bibinfo {author} {\bibfnamefont {L.}~\bibnamefont {Martin}},
  \bibinfo {author} {\bibfnamefont {J.~P.}\ \bibnamefont {Zou}}, \bibinfo
  {author} {\bibfnamefont {F.}~\bibnamefont {Mathieu}}, \bibinfo {author}
  {\bibfnamefont {P.}~\bibnamefont {Audebert}}, \bibinfo {author}
  {\bibfnamefont {P.}~\bibnamefont {Georges}},\ and\ \bibinfo {author}
  {\bibfnamefont {F.}~\bibnamefont {Druon}},\ }\bibfield  {title} {\emph
  {\bibinfo {title} {High-contrast 10 fs OPCPA-based front end for multi-PW
  laser chains}},\ }\href {https://doi.org/10.1364/OL.42.003530} {\bibfield
  {journal} {\bibinfo  {journal} {Opt. Lett.}\ }\textbf {\bibinfo {volume}
  {42}},\ \bibinfo {pages} {3530--3533} (\bibinfo {year} {2017})}\BibitemShut
  {NoStop}%
\bibitem [{\citenamefont {Papadopoulos}\ \emph {et~al.}(2016)\citenamefont
  {Papadopoulos}, \citenamefont {Zou}, \citenamefont {Le~Blanc}, \citenamefont
  {Chériaux}, \citenamefont {Georges}, \citenamefont {Druon}, \citenamefont
  {Mennerat}, \citenamefont {Ramirez}, \citenamefont {Martin}, \citenamefont
  {Fréneaux},\ and\ \citenamefont {et~al.}}]{papadopoulos_2016}%
  \BibitemOpen
  \bibfield  {author} {\bibinfo {author} {\bibfnamefont {D.}~\bibnamefont
  {Papadopoulos}}, \bibinfo {author} {\bibfnamefont {J.}~\bibnamefont {Zou}},
  \bibinfo {author} {\bibfnamefont {C.}~\bibnamefont {Le~Blanc}}, \bibinfo
  {author} {\bibfnamefont {G.}~\bibnamefont {Chériaux}}, \bibinfo {author}
  {\bibfnamefont {P.}~\bibnamefont {Georges}}, \bibinfo {author} {\bibfnamefont
  {F.}~\bibnamefont {Druon}}, \bibinfo {author} {\bibfnamefont
  {G.}~\bibnamefont {Mennerat}}, \bibinfo {author} {\bibfnamefont
  {P.}~\bibnamefont {Ramirez}}, \bibinfo {author} {\bibfnamefont
  {L.}~\bibnamefont {Martin}}, \bibinfo {author} {\bibfnamefont
  {A.}~\bibnamefont {Fréneaux}},\ and\ \bibinfo {author} {\bibnamefont
  {et~al.}},\ }\bibfield  {title} {\emph {\bibinfo {title} {The Apollon 10 PW
  laser: experimental and theoretical investigation of the temporal
  characteristics}},\ }\href {https://doi.org/10.1017/hpl.2016.34} {\bibfield
  {journal} {\bibinfo  {journal} {High Power Laser Science and Engineering}\
  }\textbf {\bibinfo {volume} {4}},\ \bibinfo {pages} {e34} (\bibinfo {year}
  {2016})}\BibitemShut {NoStop}%
\bibitem [{\citenamefont {Zou}\ \emph {et~al.}(2015)\citenamefont {Zou},
  \citenamefont {Le~Blanc}, \citenamefont {Papadopoulos}, \citenamefont
  {Chériaux}, \citenamefont {Georges}, \citenamefont {Mennerat}, \citenamefont
  {Druon}, \citenamefont {Lecherbourg}, \citenamefont {Pellegrina},
  \citenamefont {Ramirez},\ and\ \citenamefont {et~al.}}]{zou_2015}%
  \BibitemOpen
  \bibfield  {author} {\bibinfo {author} {\bibfnamefont {J.}~\bibnamefont
  {Zou}}, \bibinfo {author} {\bibfnamefont {C.}~\bibnamefont {Le~Blanc}},
  \bibinfo {author} {\bibfnamefont {D.}~\bibnamefont {Papadopoulos}}, \bibinfo
  {author} {\bibfnamefont {G.}~\bibnamefont {Chériaux}}, \bibinfo {author}
  {\bibfnamefont {P.}~\bibnamefont {Georges}}, \bibinfo {author} {\bibfnamefont
  {G.}~\bibnamefont {Mennerat}}, \bibinfo {author} {\bibfnamefont
  {F.}~\bibnamefont {Druon}}, \bibinfo {author} {\bibfnamefont
  {L.}~\bibnamefont {Lecherbourg}}, \bibinfo {author} {\bibfnamefont
  {A.}~\bibnamefont {Pellegrina}}, \bibinfo {author} {\bibfnamefont
  {P.}~\bibnamefont {Ramirez}},\ and\ \bibinfo {author} {\bibnamefont
  {et~al.}},\ }\bibfield  {title} {\emph {\bibinfo {title} {Design and current
  progress of the Apollon 10 PW project}},\ }\href
  {https://doi.org/10.1017/hpl.2014.41} {\bibfield  {journal} {\bibinfo
  {journal} {High Power Laser Science and Engineering}\ }\textbf {\bibinfo
  {volume} {3}},\ \bibinfo {pages} {e2} (\bibinfo {year} {2015})}\BibitemShut
  {NoStop}%
\bibitem [{\citenamefont {Papadopoulos}\ \emph {et~al.}(2019)\citenamefont
  {Papadopoulos}, \citenamefont {Zou}, \citenamefont {Blanc}, \citenamefont
  {Ranc}, \citenamefont {Druon}, \citenamefont {Martin}, \citenamefont
  {Fr\'{e}neaux}, \citenamefont {Beluze}, \citenamefont {Lebas}, \citenamefont
  {Chabanis}, \citenamefont {Bonnin}, \citenamefont {Accary}, \citenamefont
  {Garrec}, \citenamefont {Mathieu},\ and\ \citenamefont
  {Audebert}}]{Papadopoulos:19}%
  \BibitemOpen
  \bibfield  {author} {\bibinfo {author} {\bibfnamefont {D.~N.}\ \bibnamefont
  {Papadopoulos}}, \bibinfo {author} {\bibfnamefont {J.~P.}\ \bibnamefont
  {Zou}}, \bibinfo {author} {\bibfnamefont {C.~L.}\ \bibnamefont {Blanc}},
  \bibinfo {author} {\bibfnamefont {L.}~\bibnamefont {Ranc}}, \bibinfo {author}
  {\bibfnamefont {F.}~\bibnamefont {Druon}}, \bibinfo {author} {\bibfnamefont
  {L.}~\bibnamefont {Martin}}, \bibinfo {author} {\bibfnamefont
  {A.}~\bibnamefont {Fr\'{e}neaux}}, \bibinfo {author} {\bibfnamefont
  {A.}~\bibnamefont {Beluze}}, \bibinfo {author} {\bibfnamefont
  {N.}~\bibnamefont {Lebas}}, \bibinfo {author} {\bibfnamefont
  {M.}~\bibnamefont {Chabanis}}, \bibinfo {author} {\bibfnamefont
  {C.}~\bibnamefont {Bonnin}}, \bibinfo {author} {\bibfnamefont {J.~B.}\
  \bibnamefont {Accary}}, \bibinfo {author} {\bibfnamefont {B.~L.}\
  \bibnamefont {Garrec}}, \bibinfo {author} {\bibfnamefont {F.}~\bibnamefont
  {Mathieu}},\ and\ \bibinfo {author} {\bibfnamefont {P.}~\bibnamefont
  {Audebert}},\ }in\ \href {https://doi.org/10.1364/CLEO_SI.2019.STu3E.4}
  {\emph {\bibinfo {booktitle} {Conference on Lasers and Electro-Optics}}}\
  (\bibinfo  {publisher} {Optical Society of America},\ \bibinfo {year}
  {2019})\ p.\ \bibinfo {pages} {STu3E.4}\BibitemShut {NoStop}%
\bibitem [{\citenamefont {Papadopoulos}\ \emph {et~al.}(2021)\citenamefont
  {Papadopoulos}, \citenamefont {Zou}, \citenamefont {Blanc}, \citenamefont
  {Ranc}, \citenamefont {Druon}, \citenamefont {Martin}, \citenamefont
  {Fr\'{e}neaux}, \citenamefont {Beluze}, \citenamefont {Lebas}, \citenamefont
  {Chabanis}, \citenamefont {Garrec}, \citenamefont {Mathieu},\ and\
  \citenamefont {Audebert}}]{Papadopoulos:21}%
  \BibitemOpen
  \bibfield  {author} {\bibinfo {author} {\bibfnamefont {D.~N.}\ \bibnamefont
  {Papadopoulos}}, \bibinfo {author} {\bibfnamefont {J.~P.}\ \bibnamefont
  {Zou}}, \bibinfo {author} {\bibfnamefont {C.~L.}\ \bibnamefont {Blanc}},
  \bibinfo {author} {\bibfnamefont {L.}~\bibnamefont {Ranc}}, \bibinfo {author}
  {\bibfnamefont {F.}~\bibnamefont {Druon}}, \bibinfo {author} {\bibfnamefont
  {L.}~\bibnamefont {Martin}}, \bibinfo {author} {\bibfnamefont
  {A.}~\bibnamefont {Fr\'{e}neaux}}, \bibinfo {author} {\bibfnamefont
  {A.}~\bibnamefont {Beluze}}, \bibinfo {author} {\bibfnamefont
  {N.}~\bibnamefont {Lebas}}, \bibinfo {author} {\bibfnamefont
  {M.}~\bibnamefont {Chabanis}}, \bibinfo {author} {\bibfnamefont {B.~L.}\
  \bibnamefont {Garrec}}, \bibinfo {author} {\bibfnamefont {F.}~\bibnamefont
  {Mathieu}},\ and\ \bibinfo {author} {\bibfnamefont {P.}~\bibnamefont
  {Audebert}},\ }\bibfield  {title} {\emph {\bibinfo {title} {The Apollon
  laser: commissioning results of the 1 PW beam line}},\ }\href
  {https://v.qq.com/x/page/m3256olspwz.html} {\bibfield  {journal} {\bibinfo
  {journal} {High Power Laser Science and Engineering, April 12-16, Suzhou,
  China}\ } (\bibinfo {year} {2021})}\BibitemShut {NoStop}%
\bibitem [{\citenamefont {Dubois}\ \emph {et~al.}(2014)\citenamefont {Dubois},
  \citenamefont {Lubrano-Lavaderci}, \citenamefont {Raffestin}, \citenamefont
  {Ribolzi}, \citenamefont {Gazave}, \citenamefont {Compant La~Fontaine},
  \citenamefont {d'Humi\`{e}res}, \citenamefont {Hulin}, \citenamefont
  {Nicola\"{i}}, \citenamefont {Poy\'{e}},\ and\ \citenamefont
  {T.}}]{Dubois_PRE}%
  \BibitemOpen
  \bibfield  {author} {\bibinfo {author} {\bibfnamefont {J.-L.}\ \bibnamefont
  {Dubois}}, \bibinfo {author} {\bibfnamefont {F.}~\bibnamefont
  {Lubrano-Lavaderci}}, \bibinfo {author} {\bibfnamefont {D.}~\bibnamefont
  {Raffestin}}, \bibinfo {author} {\bibfnamefont {J.}~\bibnamefont {Ribolzi}},
  \bibinfo {author} {\bibfnamefont {J.}~\bibnamefont {Gazave}}, \bibinfo
  {author} {\bibfnamefont {A.}~\bibnamefont {Compant La~Fontaine}}, \bibinfo
  {author} {\bibfnamefont {E.}~\bibnamefont {d'Humi\`{e}res}}, \bibinfo
  {author} {\bibfnamefont {S.}~\bibnamefont {Hulin}}, \bibinfo {author}
  {\bibfnamefont {P.}~\bibnamefont {Nicola\"{i}}}, \bibinfo {author}
  {\bibfnamefont {A.}~\bibnamefont {Poy\'{e}}},\ and\ \bibinfo {author}
  {\bibfnamefont {T.~V.}\ \bibnamefont {T.}},\ }\bibfield  {title} {\emph
  {\bibinfo {title} {Target charging in short-pulse-laser-plasma
  experiments}},\ }\href {https://doi.org/10.1103/PhysRevE.89.013102}
  {\bibfield  {journal} {\bibinfo  {journal} {Physical Review E}\ }\textbf
  {\bibinfo {volume} {89}},\ \bibinfo {pages} {013202} (\bibinfo {year}
  {2014})}\BibitemShut {NoStop}%
\bibitem [{\citenamefont {Consoli}\ \emph
  {et~al.}(2020{\natexlab{a}})\citenamefont {Consoli}, \citenamefont
  {Tikhonchuk}, \citenamefont {Bardon}, \citenamefont {Bradford}, \citenamefont
  {Carroll}, \citenamefont {Cikhardt}, \citenamefont {Cipriani}, \citenamefont
  {Clarke}, \citenamefont {Cowan}, \citenamefont {Danson}, \citenamefont
  {De~Angelis}, \citenamefont {De~Marco}, \citenamefont {Dubois}, \citenamefont
  {Etchessahar}, \citenamefont {Laso~Garcia}, \citenamefont {Hillier},
  \citenamefont {Honsa}, \citenamefont {Jiang}, \citenamefont {Kmetik},
  \citenamefont {Kr\'{a}sa}, \citenamefont {Li}, \citenamefont {Lubrano},
  \citenamefont {McKenna}, \citenamefont {Metzkes-Ng}, \citenamefont {Poy\'e},
  \citenamefont {Prencipe}, \citenamefont {R\c{a}czka}, \citenamefont {Smith},
  \citenamefont {Vrana}, \citenamefont {Woolsey}, \citenamefont {Zemaityte},
  \citenamefont {Zhang}, \citenamefont {Zhang}, \citenamefont {Zielbauer},\
  and\ \citenamefont {Neely}}]{Consoli_HPLSE}%
  \BibitemOpen
  \bibfield  {author} {\bibinfo {author} {\bibfnamefont {F.}~\bibnamefont
  {Consoli}}, \bibinfo {author} {\bibfnamefont {V.~T.}\ \bibnamefont
  {Tikhonchuk}}, \bibinfo {author} {\bibfnamefont {M.}~\bibnamefont {Bardon}},
  \bibinfo {author} {\bibfnamefont {P.}~\bibnamefont {Bradford}}, \bibinfo
  {author} {\bibfnamefont {D.}~\bibnamefont {Carroll}}, \bibinfo {author}
  {\bibfnamefont {J.}~\bibnamefont {Cikhardt}}, \bibinfo {author}
  {\bibfnamefont {M.}~\bibnamefont {Cipriani}}, \bibinfo {author}
  {\bibfnamefont {R.~J.}\ \bibnamefont {Clarke}}, \bibinfo {author}
  {\bibfnamefont {T.}~\bibnamefont {Cowan}}, \bibinfo {author} {\bibfnamefont
  {C.~N.}\ \bibnamefont {Danson}}, \bibinfo {author} {\bibfnamefont
  {R.}~\bibnamefont {De~Angelis}}, \bibinfo {author} {\bibfnamefont
  {M.}~\bibnamefont {De~Marco}}, \bibinfo {author} {\bibfnamefont {J.-L.}\
  \bibnamefont {Dubois}}, \bibinfo {author} {\bibfnamefont {B.}~\bibnamefont
  {Etchessahar}}, \bibinfo {author} {\bibfnamefont {A.}~\bibnamefont
  {Laso~Garcia}}, \bibinfo {author} {\bibfnamefont {D.~I.}\ \bibnamefont
  {Hillier}}, \bibinfo {author} {\bibfnamefont {A.}~\bibnamefont {Honsa}},
  \bibinfo {author} {\bibfnamefont {W.}~\bibnamefont {Jiang}}, \bibinfo
  {author} {\bibfnamefont {V.}~\bibnamefont {Kmetik}}, \bibinfo {author}
  {\bibfnamefont {J.}~\bibnamefont {Kr\'{a}sa}}, \bibinfo {author}
  {\bibfnamefont {Y.}~\bibnamefont {Li}}, \bibinfo {author} {\bibfnamefont
  {F.}~\bibnamefont {Lubrano}}, \bibinfo {author} {\bibfnamefont
  {P.}~\bibnamefont {McKenna}}, \bibinfo {author} {\bibfnamefont
  {J.}~\bibnamefont {Metzkes-Ng}}, \bibinfo {author} {\bibfnamefont
  {A.}~\bibnamefont {Poy\'e}}, \bibinfo {author} {\bibfnamefont
  {I.}~\bibnamefont {Prencipe}}, \bibinfo {author} {\bibfnamefont
  {P.}~\bibnamefont {R\c{a}czka}}, \bibinfo {author} {\bibfnamefont {R.~A.}\
  \bibnamefont {Smith}}, \bibinfo {author} {\bibfnamefont {R.}~\bibnamefont
  {Vrana}}, \bibinfo {author} {\bibfnamefont {N.~C.}\ \bibnamefont {Woolsey}},
  \bibinfo {author} {\bibfnamefont {E.}~\bibnamefont {Zemaityte}}, \bibinfo
  {author} {\bibfnamefont {Y.}~\bibnamefont {Zhang}}, \bibinfo {author}
  {\bibfnamefont {Z.}~\bibnamefont {Zhang}}, \bibinfo {author} {\bibfnamefont
  {B.}~\bibnamefont {Zielbauer}},\ and\ \bibinfo {author} {\bibfnamefont
  {D.}~\bibnamefont {Neely}},\ }\bibfield  {title} {\emph {\bibinfo {title}
  {Laser produced electromagnetic pulses: generation, detection and
  mitigation}},\ }\href {https://doi.org/10.1017/hpl.2020.13} {\bibfield
  {journal} {\bibinfo  {journal} {High Power Laser Science and Engineering}\
  }\textbf {\bibinfo {volume} {8}},\ \bibinfo {pages} {e22} (\bibinfo {year}
  {2020}{\natexlab{a}})}\BibitemShut {NoStop}%
\bibitem [{\citenamefont {Consoli}\ \emph
  {et~al.}(2020{\natexlab{b}})\citenamefont {Consoli}, \citenamefont
  {Andreoli}, \citenamefont {Cipriani}, \citenamefont {Cristofari},
  \citenamefont {De~Angelis}, \citenamefont {Di~Giorgio}, \citenamefont
  {Duvillaret}, \citenamefont {Kr\'{a}sa}, \citenamefont {Neely}, \citenamefont
  {Salvadori}, \citenamefont {Scisciò}, \citenamefont {Smith},\ and\
  \citenamefont {Tikhonchuk}}]{Consoli_Phtransa}%
  \BibitemOpen
  \bibfield  {author} {\bibinfo {author} {\bibfnamefont {F.}~\bibnamefont
  {Consoli}}, \bibinfo {author} {\bibfnamefont {P.~L.}\ \bibnamefont
  {Andreoli}}, \bibinfo {author} {\bibfnamefont {M.}~\bibnamefont {Cipriani}},
  \bibinfo {author} {\bibfnamefont {G.}~\bibnamefont {Cristofari}}, \bibinfo
  {author} {\bibfnamefont {R.}~\bibnamefont {De~Angelis}}, \bibinfo {author}
  {\bibfnamefont {G.}~\bibnamefont {Di~Giorgio}}, \bibinfo {author}
  {\bibfnamefont {L.}~\bibnamefont {Duvillaret}}, \bibinfo {author}
  {\bibfnamefont {J.}~\bibnamefont {Kr\'{a}sa}}, \bibinfo {author}
  {\bibfnamefont {D.}~\bibnamefont {Neely}}, \bibinfo {author} {\bibfnamefont
  {M.}~\bibnamefont {Salvadori}}, \bibinfo {author} {\bibfnamefont
  {M.}~\bibnamefont {Scisciò}}, \bibinfo {author} {\bibfnamefont {R.~A.}\
  \bibnamefont {Smith}},\ and\ \bibinfo {author} {\bibfnamefont {V.~T.}\
  \bibnamefont {Tikhonchuk}},\ }\bibfield  {title} {\emph {\bibinfo {title}
  {Sources and space–time distribution of the electromagnetic pulses in
  experiments on inertial confinement fusion and laser–plasma
  acceleration}},\ }\href {https://doi.org/10.1098/rsta.2020.0022} {\bibfield
  {journal} {\bibinfo  {journal} {Philosophical Transactions of the Royal
  Society A - Mathematical, Physical and Engineering Science}\ }\textbf
  {\bibinfo {volume} {379}},\ \bibinfo {pages} {20200022} (\bibinfo {year}
  {2020}{\natexlab{b}})}\BibitemShut {NoStop}%
\bibitem [{\citenamefont {Bolton}\ \emph {et~al.}(2014)\citenamefont {Bolton},
  \citenamefont {Borghesi}, \citenamefont {Brenner}, \citenamefont {Carroll},
  \citenamefont {{De Martinis}}, \citenamefont {Fiorini}, \citenamefont
  {Flacco}, \citenamefont {Floquet}, \citenamefont {Fuchs}, \citenamefont
  {Gallegos}, \citenamefont {Giove}, \citenamefont {Green}, \citenamefont
  {Green}, \citenamefont {Jones}, \citenamefont {Kirby}, \citenamefont
  {McKenna}, \citenamefont {Neely}, \citenamefont {Nuesslin}, \citenamefont
  {Prasad}, \citenamefont {Reinhardt}, \citenamefont {Roth}, \citenamefont
  {Schramm}, \citenamefont {Scott}, \citenamefont {Ter-Avetisyan},
  \citenamefont {Tolley}, \citenamefont {Turchetti},\ and\ \citenamefont
  {Wilkens}}]{bolton2014}%
  \BibitemOpen
  \bibfield  {author} {\bibinfo {author} {\bibfnamefont {P.}~\bibnamefont
  {Bolton}}, \bibinfo {author} {\bibfnamefont {M.}~\bibnamefont {Borghesi}},
  \bibinfo {author} {\bibfnamefont {C.}~\bibnamefont {Brenner}}, \bibinfo
  {author} {\bibfnamefont {D.}~\bibnamefont {Carroll}}, \bibinfo {author}
  {\bibfnamefont {C.}~\bibnamefont {{De Martinis}}}, \bibinfo {author}
  {\bibfnamefont {F.}~\bibnamefont {Fiorini}}, \bibinfo {author} {\bibfnamefont
  {A.}~\bibnamefont {Flacco}}, \bibinfo {author} {\bibfnamefont
  {V.}~\bibnamefont {Floquet}}, \bibinfo {author} {\bibfnamefont
  {J.}~\bibnamefont {Fuchs}}, \bibinfo {author} {\bibfnamefont
  {P.}~\bibnamefont {Gallegos}}, \bibinfo {author} {\bibfnamefont
  {D.}~\bibnamefont {Giove}}, \bibinfo {author} {\bibfnamefont
  {J.}~\bibnamefont {Green}}, \bibinfo {author} {\bibfnamefont
  {S.}~\bibnamefont {Green}}, \bibinfo {author} {\bibfnamefont
  {B.}~\bibnamefont {Jones}}, \bibinfo {author} {\bibfnamefont
  {D.}~\bibnamefont {Kirby}}, \bibinfo {author} {\bibfnamefont
  {P.}~\bibnamefont {McKenna}}, \bibinfo {author} {\bibfnamefont
  {D.}~\bibnamefont {Neely}}, \bibinfo {author} {\bibfnamefont
  {F.}~\bibnamefont {Nuesslin}}, \bibinfo {author} {\bibfnamefont
  {R.}~\bibnamefont {Prasad}}, \bibinfo {author} {\bibfnamefont
  {S.}~\bibnamefont {Reinhardt}}, \bibinfo {author} {\bibfnamefont
  {M.}~\bibnamefont {Roth}}, \bibinfo {author} {\bibfnamefont {U.}~\bibnamefont
  {Schramm}}, \bibinfo {author} {\bibfnamefont {G.}~\bibnamefont {Scott}},
  \bibinfo {author} {\bibfnamefont {S.}~\bibnamefont {Ter-Avetisyan}}, \bibinfo
  {author} {\bibfnamefont {M.}~\bibnamefont {Tolley}}, \bibinfo {author}
  {\bibfnamefont {G.}~\bibnamefont {Turchetti}},\ and\ \bibinfo {author}
  {\bibfnamefont {J.}~\bibnamefont {Wilkens}},\ }\bibfield  {title} {\emph
  {\bibinfo {title} {Instrumentation for diagnostics and control of
  laser-accelerated proton (ion) beams}},\ }\href
  {https://doi.org/https://doi.org/10.1016/j.ejmp.2013.09.002} {\bibfield
  {journal} {\bibinfo  {journal} {Physica Medica}\ }\textbf {\bibinfo {volume}
  {30}},\ \bibinfo {pages} {255--270} (\bibinfo {year} {2014})}\BibitemShut
  {NoStop}%
\bibitem [{\citenamefont {Ranc}\ \emph {et~al.}(2020)\citenamefont {Ranc},
  \citenamefont {Blanc}, \citenamefont {Lebas}, \citenamefont {Martin},
  \citenamefont {Zou}, \citenamefont {Mathieu}, \citenamefont {Radier},
  \citenamefont {Ricaud}, \citenamefont {Druon},\ and\ \citenamefont
  {Papadopoulos}}]{Ranc:20}%
  \BibitemOpen
  \bibfield  {author} {\bibinfo {author} {\bibfnamefont {L.}~\bibnamefont
  {Ranc}}, \bibinfo {author} {\bibfnamefont {C.~L.}\ \bibnamefont {Blanc}},
  \bibinfo {author} {\bibfnamefont {N.}~\bibnamefont {Lebas}}, \bibinfo
  {author} {\bibfnamefont {L.}~\bibnamefont {Martin}}, \bibinfo {author}
  {\bibfnamefont {J.-P.}\ \bibnamefont {Zou}}, \bibinfo {author} {\bibfnamefont
  {F.}~\bibnamefont {Mathieu}}, \bibinfo {author} {\bibfnamefont
  {C.}~\bibnamefont {Radier}}, \bibinfo {author} {\bibfnamefont
  {S.}~\bibnamefont {Ricaud}}, \bibinfo {author} {\bibfnamefont
  {F.}~\bibnamefont {Druon}},\ and\ \bibinfo {author} {\bibfnamefont
  {D.}~\bibnamefont {Papadopoulos}},\ }\bibfield  {title} {\emph {\bibinfo
  {title} {Improvement in the temporal contrast in the tens of ps range of the
  multi-PW Apollon laser front-end}},\ }\href
  {https://doi.org/10.1364/OL.401272} {\bibfield  {journal} {\bibinfo
  {journal} {Opt. Lett.}\ }\textbf {\bibinfo {volume} {45}},\ \bibinfo {pages}
  {4599--4602} (\bibinfo {year} {2020})}\BibitemShut {NoStop}%
\bibitem [{\citenamefont {Popescu}\ \emph {et~al.}(2005)\citenamefont
  {Popescu}, \citenamefont {Baton}, \citenamefont {Amiranoff}, \citenamefont
  {Rousseaux}, \citenamefont {Le~Gloahec}, \citenamefont {Santos},
  \citenamefont {Gremillet}, \citenamefont {Koenig}, \citenamefont
  {Martinolli}, \citenamefont {Hall}, \citenamefont {Adam}, \citenamefont
  {Heron},\ and\ \citenamefont {Batani}}]{doi:10.1063/1.1927328}%
  \BibitemOpen
  \bibfield  {author} {\bibinfo {author} {\bibfnamefont {H.}~\bibnamefont
  {Popescu}}, \bibinfo {author} {\bibfnamefont {S.~D.}\ \bibnamefont {Baton}},
  \bibinfo {author} {\bibfnamefont {F.}~\bibnamefont {Amiranoff}}, \bibinfo
  {author} {\bibfnamefont {C.}~\bibnamefont {Rousseaux}}, \bibinfo {author}
  {\bibfnamefont {M.~R.}\ \bibnamefont {Le~Gloahec}}, \bibinfo {author}
  {\bibfnamefont {J.~J.}\ \bibnamefont {Santos}}, \bibinfo {author}
  {\bibfnamefont {L.}~\bibnamefont {Gremillet}}, \bibinfo {author}
  {\bibfnamefont {M.}~\bibnamefont {Koenig}}, \bibinfo {author} {\bibfnamefont
  {E.}~\bibnamefont {Martinolli}}, \bibinfo {author} {\bibfnamefont
  {T.}~\bibnamefont {Hall}}, \bibinfo {author} {\bibfnamefont {J.~C.}\
  \bibnamefont {Adam}}, \bibinfo {author} {\bibfnamefont {A.}~\bibnamefont
  {Heron}},\ and\ \bibinfo {author} {\bibfnamefont {D.}~\bibnamefont
  {Batani}},\ }\bibfield  {title} {\emph {\bibinfo {title} {Subfemtosecond,
  coherent, relativistic, and ballistic electron bunches generated at
  $\omega_0$ and 2$\omega_0$ in high intensity laser-matter interaction}},\
  }\href {https://doi.org/10.1063/1.1927328} {\bibfield  {journal} {\bibinfo
  {journal} {Physics of Plasmas}\ }\textbf {\bibinfo {volume} {12}},\ \bibinfo
  {pages} {063106} (\bibinfo {year} {2005})}\BibitemShut {NoStop}%
\bibitem [{\citenamefont {Do}\ \emph {et~al.}(2018)\citenamefont {Do},
  \citenamefont {Briat}, \citenamefont {Baton}, \citenamefont {Krumrey},
  \citenamefont {Lecherbourg}, \citenamefont {Loupias}, \citenamefont {Pérez},
  \citenamefont {Renaudin}, \citenamefont {Rubbelynck},\ and\ \citenamefont
  {Troussel}}]{doi:10.1063/1.5042069}%
  \BibitemOpen
  \bibfield  {author} {\bibinfo {author} {\bibfnamefont {A.}~\bibnamefont
  {Do}}, \bibinfo {author} {\bibfnamefont {M.}~\bibnamefont {Briat}}, \bibinfo
  {author} {\bibfnamefont {S.~D.}\ \bibnamefont {Baton}}, \bibinfo {author}
  {\bibfnamefont {M.}~\bibnamefont {Krumrey}}, \bibinfo {author} {\bibfnamefont
  {L.}~\bibnamefont {Lecherbourg}}, \bibinfo {author} {\bibfnamefont
  {B.}~\bibnamefont {Loupias}}, \bibinfo {author} {\bibfnamefont
  {F.}~\bibnamefont {Pérez}}, \bibinfo {author} {\bibfnamefont
  {P.}~\bibnamefont {Renaudin}}, \bibinfo {author} {\bibfnamefont
  {C.}~\bibnamefont {Rubbelynck}},\ and\ \bibinfo {author} {\bibfnamefont
  {P.}~\bibnamefont {Troussel}},\ }\bibfield  {title} {\emph {\bibinfo {title}
  {Two-channel high-resolution quasi-monochromatic X-ray imager for Al and Ti
  plasma}},\ }\href {https://doi.org/10.1063/1.5042069} {\bibfield  {journal}
  {\bibinfo  {journal} {Review of Scientific Instruments}\ }\textbf {\bibinfo
  {volume} {89}},\ \bibinfo {pages} {113702} (\bibinfo {year}
  {2018})}\BibitemShut {NoStop}%
\bibitem [{\citenamefont {Renaudin}\ \emph {et~al.}(2020)\citenamefont
  {Renaudin}, \citenamefont {Duthoit}, \citenamefont {Baton}, \citenamefont
  {Blancard}, \citenamefont {Chaleil}, \citenamefont {Coss{\'e}}, \citenamefont
  {Faussurier}, \citenamefont {Gremillet}, \citenamefont {Lecherbourg},
  \citenamefont {Loupias},\ and\ \citenamefont
  {Perez}}]{renaudin:hal-02456723}%
  \BibitemOpen
  \bibfield  {author} {\bibinfo {author} {\bibfnamefont {P.}~\bibnamefont
  {Renaudin}}, \bibinfo {author} {\bibfnamefont {L.}~\bibnamefont {Duthoit}},
  \bibinfo {author} {\bibfnamefont {S.}~\bibnamefont {Baton}}, \bibinfo
  {author} {\bibfnamefont {C.}~\bibnamefont {Blancard}}, \bibinfo {author}
  {\bibfnamefont {A.}~\bibnamefont {Chaleil}}, \bibinfo {author} {\bibfnamefont
  {P.}~\bibnamefont {Coss{\'e}}}, \bibinfo {author} {\bibfnamefont
  {G.}~\bibnamefont {Faussurier}}, \bibinfo {author} {\bibfnamefont
  {L.}~\bibnamefont {Gremillet}}, \bibinfo {author} {\bibfnamefont
  {L.}~\bibnamefont {Lecherbourg}}, \bibinfo {author} {\bibfnamefont
  {B.}~\bibnamefont {Loupias}},\ and\ \bibinfo {author} {\bibfnamefont
  {F.}~\bibnamefont {Perez}},\ }\bibfield  {title} {\emph {\bibinfo {title}
  {{Radiative cooling of an Al plasma in an Al-Ti mixture heated by an
  ultraintense laser pulse}}},\ }\href
  {https://hal.archives-ouvertes.fr/hal-02456723} {\bibfield  {journal}
  {\bibinfo  {journal} {{High Energy Density Physics}}\ } (\bibinfo {year}
  {2020})}\BibitemShut {NoStop}%
\bibitem [{\citenamefont {Faenov}\ \emph {et~al.}(1994)\citenamefont {Faenov},
  \citenamefont {Pikuz}, \citenamefont {Erko}, \citenamefont {Bryunetkin},
  \citenamefont {Dyakin}, \citenamefont {Ivanenkov}, \citenamefont {Mingaleev},
  \citenamefont {Pikuz}, \citenamefont {Romanova},\ and\ \citenamefont
  {Shelkovenko}}]{Faenov_1994}%
  \BibitemOpen
  \bibfield  {author} {\bibinfo {author} {\bibfnamefont {A.~Y.}\ \bibnamefont
  {Faenov}}, \bibinfo {author} {\bibfnamefont {S.~A.}\ \bibnamefont {Pikuz}},
  \bibinfo {author} {\bibfnamefont {A.~I.}\ \bibnamefont {Erko}}, \bibinfo
  {author} {\bibfnamefont {B.~A.}\ \bibnamefont {Bryunetkin}}, \bibinfo
  {author} {\bibfnamefont {V.~M.}\ \bibnamefont {Dyakin}}, \bibinfo {author}
  {\bibfnamefont {G.~V.}\ \bibnamefont {Ivanenkov}}, \bibinfo {author}
  {\bibfnamefont {A.~R.}\ \bibnamefont {Mingaleev}}, \bibinfo {author}
  {\bibfnamefont {T.~A.}\ \bibnamefont {Pikuz}}, \bibinfo {author}
  {\bibfnamefont {V.~M.}\ \bibnamefont {Romanova}},\ and\ \bibinfo {author}
  {\bibfnamefont {T.~A.}\ \bibnamefont {Shelkovenko}},\ }\bibfield  {title}
  {\emph {\bibinfo {title} {High-performance x-ray spectroscopic devices for
  plasma microsources investigations}},\ }\href
  {https://doi.org/10.1088/0031-8949/50/4/003} {\bibfield  {journal} {\bibinfo
  {journal} {Physica Scripta}\ }\textbf {\bibinfo {volume} {50}},\ \bibinfo
  {pages} {333--338} (\bibinfo {year} {1994})}\BibitemShut {NoStop}%
\bibitem [{\citenamefont {Andiel}\ \emph {et~al.}(2002)\citenamefont {Andiel},
  \citenamefont {Eidmann}, \citenamefont {Hakel}, \citenamefont {Mancini},
  \citenamefont {Junkel-Vives}, \citenamefont {Abdallah},\ and\ \citenamefont
  {Witte}}]{Andiel2002}%
  \BibitemOpen
  \bibfield  {author} {\bibinfo {author} {\bibfnamefont {U.}~\bibnamefont
  {Andiel}}, \bibinfo {author} {\bibfnamefont {K.}~\bibnamefont {Eidmann}},
  \bibinfo {author} {\bibfnamefont {P.}~\bibnamefont {Hakel}}, \bibinfo
  {author} {\bibfnamefont {R.~C.}\ \bibnamefont {Mancini}}, \bibinfo {author}
  {\bibfnamefont {G.~C.}\ \bibnamefont {Junkel-Vives}}, \bibinfo {author}
  {\bibfnamefont {J.}~\bibnamefont {Abdallah}},\ and\ \bibinfo {author}
  {\bibfnamefont {K.}~\bibnamefont {Witte}},\ }\bibfield  {title} {\emph
  {\bibinfo {title} {Demonstration of aluminum K-shell line shifts in
  isochorically heated targets driven by ultrashort laser pulses}},\ }\href
  {https://doi.org/10.1209/epl/i2002-00297-5} {\bibfield  {journal} {\bibinfo
  {journal} {Europhysics Letters ({EPL})}\ }\textbf {\bibinfo {volume} {60}},\
  \bibinfo {pages} {861--867} (\bibinfo {year} {2002})}\BibitemShut {NoStop}%
\bibitem [{\citenamefont {Alkhimova}\ \emph {et~al.}(2017)\citenamefont
  {Alkhimova}, \citenamefont {Faenov}, \citenamefont {Skobelev}, \citenamefont
  {Pikuz}, \citenamefont {Nishiuchi}, \citenamefont {Sakaki}, \citenamefont
  {Pirozhkov}, \citenamefont {Sagisaka}, \citenamefont {Dover}, \citenamefont
  {Kondo}, \citenamefont {Ogura}, \citenamefont {Fukuda}, \citenamefont
  {Kiriyama}, \citenamefont {Nishitani}, \citenamefont {Miyahara},
  \citenamefont {Watanabe}, \citenamefont {Pikuz}, \citenamefont {Kando},
  \citenamefont {Kodama},\ and\ \citenamefont {Kondo}}]{Alkhimova2017}%
  \BibitemOpen
  \bibfield  {author} {\bibinfo {author} {\bibfnamefont {M.~A.}\ \bibnamefont
  {Alkhimova}}, \bibinfo {author} {\bibfnamefont {A.~Y.}\ \bibnamefont
  {Faenov}}, \bibinfo {author} {\bibfnamefont {I.~Y.}\ \bibnamefont
  {Skobelev}}, \bibinfo {author} {\bibfnamefont {T.~A.}\ \bibnamefont {Pikuz}},
  \bibinfo {author} {\bibfnamefont {M.}~\bibnamefont {Nishiuchi}}, \bibinfo
  {author} {\bibfnamefont {H.}~\bibnamefont {Sakaki}}, \bibinfo {author}
  {\bibfnamefont {A.~S.}\ \bibnamefont {Pirozhkov}}, \bibinfo {author}
  {\bibfnamefont {A.}~\bibnamefont {Sagisaka}}, \bibinfo {author}
  {\bibfnamefont {N.~P.}\ \bibnamefont {Dover}}, \bibinfo {author}
  {\bibfnamefont {K.}~\bibnamefont {Kondo}}, \bibinfo {author} {\bibfnamefont
  {K.}~\bibnamefont {Ogura}}, \bibinfo {author} {\bibfnamefont
  {Y.}~\bibnamefont {Fukuda}}, \bibinfo {author} {\bibfnamefont
  {H.}~\bibnamefont {Kiriyama}}, \bibinfo {author} {\bibfnamefont
  {K.}~\bibnamefont {Nishitani}}, \bibinfo {author} {\bibfnamefont
  {T.}~\bibnamefont {Miyahara}}, \bibinfo {author} {\bibfnamefont
  {Y.}~\bibnamefont {Watanabe}}, \bibinfo {author} {\bibfnamefont {S.~A.}\
  \bibnamefont {Pikuz}}, \bibinfo {author} {\bibfnamefont {M.}~\bibnamefont
  {Kando}}, \bibinfo {author} {\bibfnamefont {R.}~\bibnamefont {Kodama}},\ and\
  \bibinfo {author} {\bibfnamefont {K.}~\bibnamefont {Kondo}},\ }\bibfield
  {title} {\emph {\bibinfo {title} {High resolution X-ray spectra of stainless
  steel foils irradiated by femtosecond laser pulses with ultra-relativistic
  intensities}},\ }\href {https://doi.org/10.1364/OE.25.029501} {\bibfield
  {journal} {\bibinfo  {journal} {Opt. Express}\ }\textbf {\bibinfo {volume}
  {25}},\ \bibinfo {pages} {29501--29511} (\bibinfo {year} {2017})}\BibitemShut
  {NoStop}%
\bibitem [{\citenamefont {Chopineau}\ \emph {et~al.}(2019)\citenamefont
  {Chopineau}, \citenamefont {Leblanc}, \citenamefont {Blaclard}, \citenamefont
  {Denoeud}, \citenamefont {Th\'evenet}, \citenamefont {Vay}, \citenamefont
  {Bonnaud}, \citenamefont {Martin}, \citenamefont {Vincenti},\ and\
  \citenamefont {Qu\'er\'e}}]{PhysRevX.9.011050}%
  \BibitemOpen
  \bibfield  {author} {\bibinfo {author} {\bibfnamefont {L.}~\bibnamefont
  {Chopineau}}, \bibinfo {author} {\bibfnamefont {A.}~\bibnamefont {Leblanc}},
  \bibinfo {author} {\bibfnamefont {G.}~\bibnamefont {Blaclard}}, \bibinfo
  {author} {\bibfnamefont {A.}~\bibnamefont {Denoeud}}, \bibinfo {author}
  {\bibfnamefont {M.}~\bibnamefont {Th\'evenet}}, \bibinfo {author}
  {\bibfnamefont {J.-L.}\ \bibnamefont {Vay}}, \bibinfo {author} {\bibfnamefont
  {G.}~\bibnamefont {Bonnaud}}, \bibinfo {author} {\bibfnamefont
  {P.}~\bibnamefont {Martin}}, \bibinfo {author} {\bibfnamefont
  {H.}~\bibnamefont {Vincenti}},\ and\ \bibinfo {author} {\bibfnamefont
  {F.}~\bibnamefont {Qu\'er\'e}},\ }\bibfield  {title} {\emph {\bibinfo {title}
  {Identification of Coupling Mechanisms between Ultraintense Laser Light and
  Dense Plasmas}},\ }\href {https://doi.org/10.1103/PhysRevX.9.011050}
  {\bibfield  {journal} {\bibinfo  {journal} {Phys. Rev. X}\ }\textbf {\bibinfo
  {volume} {9}},\ \bibinfo {pages} {011050} (\bibinfo {year}
  {2019})}\BibitemShut {NoStop}%
\bibitem [{\citenamefont {Kahaly}\ \emph {et~al.}(2013)\citenamefont {Kahaly},
  \citenamefont {Monchoc\'e}, \citenamefont {Vincenti}, \citenamefont
  {Dzelzainis}, \citenamefont {Dromey}, \citenamefont {Zepf}, \citenamefont
  {Martin},\ and\ \citenamefont {Qu\'er\'e}}]{PhysRevLett.110.175001}%
  \BibitemOpen
  \bibfield  {author} {\bibinfo {author} {\bibfnamefont {S.}~\bibnamefont
  {Kahaly}}, \bibinfo {author} {\bibfnamefont {S.}~\bibnamefont {Monchoc\'e}},
  \bibinfo {author} {\bibfnamefont {H.}~\bibnamefont {Vincenti}}, \bibinfo
  {author} {\bibfnamefont {T.}~\bibnamefont {Dzelzainis}}, \bibinfo {author}
  {\bibfnamefont {B.}~\bibnamefont {Dromey}}, \bibinfo {author} {\bibfnamefont
  {M.}~\bibnamefont {Zepf}}, \bibinfo {author} {\bibfnamefont {P.}~\bibnamefont
  {Martin}},\ and\ \bibinfo {author} {\bibfnamefont {F.}~\bibnamefont
  {Qu\'er\'e}},\ }\bibfield  {title} {\emph {\bibinfo {title} {Direct
  Observation of Density-Gradient Effects in Harmonic Generation from Plasma
  Mirrors}},\ }\href {https://doi.org/10.1103/PhysRevLett.110.175001}
  {\bibfield  {journal} {\bibinfo  {journal} {Phys. Rev. Lett.}\ }\textbf
  {\bibinfo {volume} {110}},\ \bibinfo {pages} {175001} (\bibinfo {year}
  {2013})}\BibitemShut {NoStop}%
\bibitem [{\citenamefont {Salamin}\ \emph {et~al.}(2006)\citenamefont
  {Salamin}, \citenamefont {Hu}, \citenamefont {Hatsagortsyan},\ and\
  \citenamefont {Keitel}}]{salamin2006}%
  \BibitemOpen
  \bibfield  {author} {\bibinfo {author} {\bibfnamefont {Y.~I.}\ \bibnamefont
  {Salamin}}, \bibinfo {author} {\bibfnamefont {S.}~\bibnamefont {Hu}},
  \bibinfo {author} {\bibfnamefont {K.~Z.}\ \bibnamefont {Hatsagortsyan}},\
  and\ \bibinfo {author} {\bibfnamefont {C.~H.}\ \bibnamefont {Keitel}},\
  }\bibfield  {title} {\emph {\bibinfo {title} {Relativistic high-power
  laser–matter interactions}},\ }\href
  {https://doi.org/10.1016/j.physrep.2006.01.002} {\bibfield  {journal}
  {\bibinfo  {journal} {Physics Reports}\ }\textbf {\bibinfo {volume} {427}},\
  \bibinfo {pages} {41--155} (\bibinfo {year} {2006})}\BibitemShut {NoStop}%
\bibitem [{\citenamefont {L\'{e}vy}\ \emph {et~al.}(2007)\citenamefont
  {L\'{e}vy}, \citenamefont {Ceccotti}, \citenamefont {D'Oliveira},
  \citenamefont {R\'{e}au}, \citenamefont {Perdrix}, \citenamefont
  {Qu\'{e}r\'{e}}, \citenamefont {Monot}, \citenamefont {Bougeard},
  \citenamefont {Lagadec}, \citenamefont {Martin}, \citenamefont {Geindre},\
  and\ \citenamefont {Audebert}}]{Levy:07}%
  \BibitemOpen
  \bibfield  {author} {\bibinfo {author} {\bibfnamefont {A.}~\bibnamefont
  {L\'{e}vy}}, \bibinfo {author} {\bibfnamefont {T.}~\bibnamefont {Ceccotti}},
  \bibinfo {author} {\bibfnamefont {P.}~\bibnamefont {D'Oliveira}}, \bibinfo
  {author} {\bibfnamefont {F.}~\bibnamefont {R\'{e}au}}, \bibinfo {author}
  {\bibfnamefont {M.}~\bibnamefont {Perdrix}}, \bibinfo {author} {\bibfnamefont
  {F.}~\bibnamefont {Qu\'{e}r\'{e}}}, \bibinfo {author} {\bibfnamefont
  {P.}~\bibnamefont {Monot}}, \bibinfo {author} {\bibfnamefont
  {M.}~\bibnamefont {Bougeard}}, \bibinfo {author} {\bibfnamefont
  {H.}~\bibnamefont {Lagadec}}, \bibinfo {author} {\bibfnamefont
  {P.}~\bibnamefont {Martin}}, \bibinfo {author} {\bibfnamefont {J.-P.}\
  \bibnamefont {Geindre}},\ and\ \bibinfo {author} {\bibfnamefont
  {P.}~\bibnamefont {Audebert}},\ }\bibfield  {title} {\emph {\bibinfo {title}
  {Double plasma mirror for ultrahigh temporal contrast ultraintense laser
  pulses}},\ }\href {https://doi.org/10.1364/OL.32.000310} {\bibfield
  {journal} {\bibinfo  {journal} {Opt. Lett.}\ }\textbf {\bibinfo {volume}
  {32}},\ \bibinfo {pages} {310--312} (\bibinfo {year} {2007})}\BibitemShut
  {NoStop}%
\bibitem [{\citenamefont {Snavely}\ \emph {et~al.}(2000)\citenamefont
  {Snavely}, \citenamefont {Key}, \citenamefont {Hatchett}, \citenamefont
  {Cowan}, \citenamefont {Roth}, \citenamefont {Phillips}, \citenamefont
  {Stoyer}, \citenamefont {Henry}, \citenamefont {Sangster}, \citenamefont
  {Singh}, \citenamefont {Wilks}, \citenamefont {MacKinnon}, \citenamefont
  {Offenberger}, \citenamefont {Pennington}, \citenamefont {Yasuike},
  \citenamefont {Langdon}, \citenamefont {Lasinski}, \citenamefont {Johnson},
  \citenamefont {Perry},\ and\ \citenamefont {Campbell}}]{PhysRevLett.85.2945}%
  \BibitemOpen
  \bibfield  {author} {\bibinfo {author} {\bibfnamefont {R.~A.}\ \bibnamefont
  {Snavely}}, \bibinfo {author} {\bibfnamefont {M.~H.}\ \bibnamefont {Key}},
  \bibinfo {author} {\bibfnamefont {S.~P.}\ \bibnamefont {Hatchett}}, \bibinfo
  {author} {\bibfnamefont {T.~E.}\ \bibnamefont {Cowan}}, \bibinfo {author}
  {\bibfnamefont {M.}~\bibnamefont {Roth}}, \bibinfo {author} {\bibfnamefont
  {T.~W.}\ \bibnamefont {Phillips}}, \bibinfo {author} {\bibfnamefont {M.~A.}\
  \bibnamefont {Stoyer}}, \bibinfo {author} {\bibfnamefont {E.~A.}\
  \bibnamefont {Henry}}, \bibinfo {author} {\bibfnamefont {T.~C.}\ \bibnamefont
  {Sangster}}, \bibinfo {author} {\bibfnamefont {M.~S.}\ \bibnamefont {Singh}},
  \bibinfo {author} {\bibfnamefont {S.~C.}\ \bibnamefont {Wilks}}, \bibinfo
  {author} {\bibfnamefont {A.}~\bibnamefont {MacKinnon}}, \bibinfo {author}
  {\bibfnamefont {A.}~\bibnamefont {Offenberger}}, \bibinfo {author}
  {\bibfnamefont {D.~M.}\ \bibnamefont {Pennington}}, \bibinfo {author}
  {\bibfnamefont {K.}~\bibnamefont {Yasuike}}, \bibinfo {author} {\bibfnamefont
  {A.~B.}\ \bibnamefont {Langdon}}, \bibinfo {author} {\bibfnamefont {B.~F.}\
  \bibnamefont {Lasinski}}, \bibinfo {author} {\bibfnamefont {J.}~\bibnamefont
  {Johnson}}, \bibinfo {author} {\bibfnamefont {M.~D.}\ \bibnamefont {Perry}},\
  and\ \bibinfo {author} {\bibfnamefont {E.~M.}\ \bibnamefont {Campbell}},\
  }\bibfield  {title} {\emph {\bibinfo {title} {Intense High-Energy Proton
  Beams from Petawatt-Laser Irradiation of Solids}},\ }\href
  {https://doi.org/10.1103/PhysRevLett.85.2945} {\bibfield  {journal} {\bibinfo
   {journal} {Phys. Rev. Lett.}\ }\textbf {\bibinfo {volume} {85}},\ \bibinfo
  {pages} {2945--2948} (\bibinfo {year} {2000})}\BibitemShut {NoStop}%
\bibitem [{\citenamefont {Thomson}(1911)}]{doi:10.1080/14786440208637024}%
  \BibitemOpen
  \bibfield  {author} {\bibinfo {author} {\bibfnamefont {S.~J.}\ \bibnamefont
  {Thomson}},\ }\bibfield  {title} {\emph {\bibinfo {title} {Rays of positive
  electricity}},\ }\href {https://doi.org/10.1080/14786440208637024} {\bibfield
   {journal} {\bibinfo  {journal} {The London, Edinburgh, and Dublin
  Philosophical Magazine and Journal of Science}\ }\textbf {\bibinfo {volume}
  {21}},\ \bibinfo {pages} {225--249} (\bibinfo {year} {1911})}\BibitemShut
  {NoStop}%
\bibitem [{\citenamefont {Fuchs}\ \emph {et~al.}(2006)\citenamefont {Fuchs},
  \citenamefont {Antici}, \citenamefont {d'Humi{\`e}res}, \citenamefont
  {Lefebvre}, \citenamefont {Borghesi}, \citenamefont {Brambrink},
  \citenamefont {Cecchetti}, \citenamefont {Kaluza}, \citenamefont {Malka},
  \citenamefont {Manclossi}, \citenamefont {Meyroneinc}, \citenamefont {Mora},
  \citenamefont {Schreiber}, \citenamefont {Toncian}, \citenamefont
  {P{\'e}pin},\ and\ \citenamefont {Audebert}}]{Fuchs2006natphys}%
  \BibitemOpen
  \bibfield  {author} {\bibinfo {author} {\bibfnamefont {J.}~\bibnamefont
  {Fuchs}}, \bibinfo {author} {\bibfnamefont {P.}~\bibnamefont {Antici}},
  \bibinfo {author} {\bibfnamefont {E.}~\bibnamefont {d'Humi{\`e}res}},
  \bibinfo {author} {\bibfnamefont {E.}~\bibnamefont {Lefebvre}}, \bibinfo
  {author} {\bibfnamefont {M.}~\bibnamefont {Borghesi}}, \bibinfo {author}
  {\bibfnamefont {E.}~\bibnamefont {Brambrink}}, \bibinfo {author}
  {\bibfnamefont {C.~A.}\ \bibnamefont {Cecchetti}}, \bibinfo {author}
  {\bibfnamefont {M.}~\bibnamefont {Kaluza}}, \bibinfo {author} {\bibfnamefont
  {V.}~\bibnamefont {Malka}}, \bibinfo {author} {\bibfnamefont
  {M.}~\bibnamefont {Manclossi}}, \bibinfo {author} {\bibfnamefont
  {S.}~\bibnamefont {Meyroneinc}}, \bibinfo {author} {\bibfnamefont
  {P.}~\bibnamefont {Mora}}, \bibinfo {author} {\bibfnamefont {J.}~\bibnamefont
  {Schreiber}}, \bibinfo {author} {\bibfnamefont {T.}~\bibnamefont {Toncian}},
  \bibinfo {author} {\bibfnamefont {H.}~\bibnamefont {P{\'e}pin}},\ and\
  \bibinfo {author} {\bibfnamefont {P.}~\bibnamefont {Audebert}},\ }\bibfield
  {title} {\emph {\bibinfo {title} {Laser-driven proton scaling laws and new
  paths towards energy increase}},\ }\href {https://doi.org/10.1038/nphys199}
  {\bibfield  {journal} {\bibinfo  {journal} {Nature Physics}\ }\textbf
  {\bibinfo {volume} {2}},\ \bibinfo {pages} {48--54} (\bibinfo {year}
  {2006})}\BibitemShut {NoStop}%
\bibitem [{\citenamefont {Ceccotti}\ \emph {et~al.}(2007)\citenamefont
  {Ceccotti}, \citenamefont {L\'evy}, \citenamefont {Popescu}, \citenamefont
  {R\'eau}, \citenamefont {D'Oliveira}, \citenamefont {Monot}, \citenamefont
  {Geindre}, \citenamefont {Lefebvre},\ and\ \citenamefont
  {Martin}}]{PhysRevLett.99.185002}%
  \BibitemOpen
  \bibfield  {author} {\bibinfo {author} {\bibfnamefont {T.}~\bibnamefont
  {Ceccotti}}, \bibinfo {author} {\bibfnamefont {A.}~\bibnamefont {L\'evy}},
  \bibinfo {author} {\bibfnamefont {H.}~\bibnamefont {Popescu}}, \bibinfo
  {author} {\bibfnamefont {F.}~\bibnamefont {R\'eau}}, \bibinfo {author}
  {\bibfnamefont {P.}~\bibnamefont {D'Oliveira}}, \bibinfo {author}
  {\bibfnamefont {P.}~\bibnamefont {Monot}}, \bibinfo {author} {\bibfnamefont
  {J.~P.}\ \bibnamefont {Geindre}}, \bibinfo {author} {\bibfnamefont
  {E.}~\bibnamefont {Lefebvre}},\ and\ \bibinfo {author} {\bibfnamefont
  {P.}~\bibnamefont {Martin}},\ }\bibfield  {title} {\emph {\bibinfo {title}
  {Proton Acceleration with High-Intensity Ultrahigh-Contrast Laser Pulses}},\
  }\href {https://doi.org/10.1103/PhysRevLett.99.185002} {\bibfield  {journal}
  {\bibinfo  {journal} {Phys. Rev. Lett.}\ }\textbf {\bibinfo {volume} {99}},\
  \bibinfo {pages} {185002} (\bibinfo {year} {2007})}\BibitemShut {NoStop}%
\bibitem [{\citenamefont {Soloviev}\ \emph {et~al.}(2017)\citenamefont
  {Soloviev}, \citenamefont {Burdonov}, \citenamefont {Chen}, \citenamefont
  {Eremeev}, \citenamefont {Korzhimanov}, \citenamefont {Pokrovskiy},
  \citenamefont {Pikuz}, \citenamefont {Revet}, \citenamefont {Sladkov},
  \citenamefont {Ginzburg}, \citenamefont {Khazanov}, \citenamefont {Kuzmin},
  \citenamefont {Osmanov}, \citenamefont {Shaikin}, \citenamefont {Shaykin},
  \citenamefont {Yakovlev}, \citenamefont {Pikuz}, \citenamefont
  {Starodubtsev},\ and\ \citenamefont {Fuchs}}]{Soloviev2017}%
  \BibitemOpen
  \bibfield  {author} {\bibinfo {author} {\bibfnamefont {A.}~\bibnamefont
  {Soloviev}}, \bibinfo {author} {\bibfnamefont {K.}~\bibnamefont {Burdonov}},
  \bibinfo {author} {\bibfnamefont {S.~N.}\ \bibnamefont {Chen}}, \bibinfo
  {author} {\bibfnamefont {A.}~\bibnamefont {Eremeev}}, \bibinfo {author}
  {\bibfnamefont {A.}~\bibnamefont {Korzhimanov}}, \bibinfo {author}
  {\bibfnamefont {G.~V.}\ \bibnamefont {Pokrovskiy}}, \bibinfo {author}
  {\bibfnamefont {T.~A.}\ \bibnamefont {Pikuz}}, \bibinfo {author}
  {\bibfnamefont {G.}~\bibnamefont {Revet}}, \bibinfo {author} {\bibfnamefont
  {A.}~\bibnamefont {Sladkov}}, \bibinfo {author} {\bibfnamefont
  {V.}~\bibnamefont {Ginzburg}}, \bibinfo {author} {\bibfnamefont
  {E.}~\bibnamefont {Khazanov}}, \bibinfo {author} {\bibfnamefont
  {A.}~\bibnamefont {Kuzmin}}, \bibinfo {author} {\bibfnamefont
  {R.}~\bibnamefont {Osmanov}}, \bibinfo {author} {\bibfnamefont
  {I.}~\bibnamefont {Shaikin}}, \bibinfo {author} {\bibfnamefont
  {A.}~\bibnamefont {Shaykin}}, \bibinfo {author} {\bibfnamefont
  {I.}~\bibnamefont {Yakovlev}}, \bibinfo {author} {\bibfnamefont
  {S.}~\bibnamefont {Pikuz}}, \bibinfo {author} {\bibfnamefont
  {M.}~\bibnamefont {Starodubtsev}},\ and\ \bibinfo {author} {\bibfnamefont
  {J.}~\bibnamefont {Fuchs}},\ }\bibfield  {title} {\emph {\bibinfo {title}
  {Experimental evidence for short-pulse laser heating of solid-density target
  to high bulk temperatures}},\ }\href
  {https://doi.org/10.1038/s41598-017-11675-2} {\bibfield  {journal} {\bibinfo
  {journal} {Scientific Reports}\ }\textbf {\bibinfo {volume} {7}},\ \bibinfo
  {pages} {12144} (\bibinfo {year} {2017})}\BibitemShut {NoStop}%
\bibitem [{\citenamefont {Wilks}\ \emph {et~al.}(2001)\citenamefont {Wilks},
  \citenamefont {Langdon}, \citenamefont {Cowan}, \citenamefont {Roth},
  \citenamefont {Singh}, \citenamefont {Hatchett}, \citenamefont {Key},
  \citenamefont {Pennington}, \citenamefont {MacKinnon},\ and\ \citenamefont
  {Snavely}}]{doi:10.1063/1.1333697}%
  \BibitemOpen
  \bibfield  {author} {\bibinfo {author} {\bibfnamefont {S.~C.}\ \bibnamefont
  {Wilks}}, \bibinfo {author} {\bibfnamefont {A.~B.}\ \bibnamefont {Langdon}},
  \bibinfo {author} {\bibfnamefont {T.~E.}\ \bibnamefont {Cowan}}, \bibinfo
  {author} {\bibfnamefont {M.}~\bibnamefont {Roth}}, \bibinfo {author}
  {\bibfnamefont {M.}~\bibnamefont {Singh}}, \bibinfo {author} {\bibfnamefont
  {S.}~\bibnamefont {Hatchett}}, \bibinfo {author} {\bibfnamefont {M.~H.}\
  \bibnamefont {Key}}, \bibinfo {author} {\bibfnamefont {D.}~\bibnamefont
  {Pennington}}, \bibinfo {author} {\bibfnamefont {A.}~\bibnamefont
  {MacKinnon}},\ and\ \bibinfo {author} {\bibfnamefont {R.~A.}\ \bibnamefont
  {Snavely}},\ }\bibfield  {title} {\emph {\bibinfo {title} {Energetic proton
  generation in ultra-intense laser–solid interactions}},\ }\href
  {https://doi.org/10.1063/1.1333697} {\bibfield  {journal} {\bibinfo
  {journal} {Physics of Plasmas}\ }\textbf {\bibinfo {volume} {8}},\ \bibinfo
  {pages} {542--549} (\bibinfo {year} {2001})}\BibitemShut {NoStop}%
\bibitem [{\citenamefont {Daido}\ \emph {et~al.}(2012)\citenamefont {Daido},
  \citenamefont {Nishiuchi},\ and\ \citenamefont {Pirozhkov}}]{Daido_2012}%
  \BibitemOpen
  \bibfield  {author} {\bibinfo {author} {\bibfnamefont {H.}~\bibnamefont
  {Daido}}, \bibinfo {author} {\bibfnamefont {M.}~\bibnamefont {Nishiuchi}},\
  and\ \bibinfo {author} {\bibfnamefont {A.~S.}\ \bibnamefont {Pirozhkov}},\
  }\bibfield  {title} {\emph {\bibinfo {title} {Review of laser-driven ion
  sources and their applications}},\ }\href
  {https://doi.org/10.1088/0034-4885/75/5/056401} {\bibfield  {journal}
  {\bibinfo  {journal} {Reports on Progress in Physics}\ }\textbf {\bibinfo
  {volume} {75}},\ \bibinfo {pages} {056401} (\bibinfo {year}
  {2012})}\BibitemShut {NoStop}%
\bibitem [{\citenamefont {Macchi}\ \emph {et~al.}(2013)\citenamefont {Macchi},
  \citenamefont {Borghesi},\ and\ \citenamefont {Passoni}}]{RevModPhys.85.751}%
  \BibitemOpen
  \bibfield  {author} {\bibinfo {author} {\bibfnamefont {A.}~\bibnamefont
  {Macchi}}, \bibinfo {author} {\bibfnamefont {M.}~\bibnamefont {Borghesi}},\
  and\ \bibinfo {author} {\bibfnamefont {M.}~\bibnamefont {Passoni}},\
  }\bibfield  {title} {\emph {\bibinfo {title} {Ion acceleration by
  superintense laser-plasma interaction}},\ }\href
  {https://doi.org/10.1103/RevModPhys.85.751} {\bibfield  {journal} {\bibinfo
  {journal} {Rev. Mod. Phys.}\ }\textbf {\bibinfo {volume} {85}},\ \bibinfo
  {pages} {751--793} (\bibinfo {year} {2013})}\BibitemShut {NoStop}%
\bibitem [{\citenamefont {Wagner}\ \emph
  {et~al.}(2016{\natexlab{a}})\citenamefont {Wagner}, \citenamefont {Deppert},
  \citenamefont {Brabetz}, \citenamefont {Fiala}, \citenamefont {Kleinschmidt},
  \citenamefont {Poth}, \citenamefont {Schanz}, \citenamefont {Tebartz},
  \citenamefont {Zielbauer}, \citenamefont {Roth}, \citenamefont {St\"ohlker},\
  and\ \citenamefont {Bagnoud}}]{PhysRevLett.116.205002}%
  \BibitemOpen
  \bibfield  {author} {\bibinfo {author} {\bibfnamefont {F.}~\bibnamefont
  {Wagner}}, \bibinfo {author} {\bibfnamefont {O.}~\bibnamefont {Deppert}},
  \bibinfo {author} {\bibfnamefont {C.}~\bibnamefont {Brabetz}}, \bibinfo
  {author} {\bibfnamefont {P.}~\bibnamefont {Fiala}}, \bibinfo {author}
  {\bibfnamefont {A.}~\bibnamefont {Kleinschmidt}}, \bibinfo {author}
  {\bibfnamefont {P.}~\bibnamefont {Poth}}, \bibinfo {author} {\bibfnamefont
  {V.~A.}\ \bibnamefont {Schanz}}, \bibinfo {author} {\bibfnamefont
  {A.}~\bibnamefont {Tebartz}}, \bibinfo {author} {\bibfnamefont
  {B.}~\bibnamefont {Zielbauer}}, \bibinfo {author} {\bibfnamefont
  {M.}~\bibnamefont {Roth}}, \bibinfo {author} {\bibfnamefont {T.}~\bibnamefont
  {St\"ohlker}},\ and\ \bibinfo {author} {\bibfnamefont {V.}~\bibnamefont
  {Bagnoud}},\ }\bibfield  {title} {\emph {\bibinfo {title} {Maximum Proton
  Energy above 85 MeV from the Relativistic Interaction of Laser Pulses with
  Micrometer Thick ${\mathrm{CH}}_{2}$ Targets}},\ }\href
  {https://doi.org/10.1103/PhysRevLett.116.205002} {\bibfield  {journal}
  {\bibinfo  {journal} {Phys. Rev. Lett.}\ }\textbf {\bibinfo {volume} {116}},\
  \bibinfo {pages} {205002} (\bibinfo {year} {2016}{\natexlab{a}})}\BibitemShut
  {NoStop}%
\bibitem [{\citenamefont {Adam}\ \emph {et~al.}(2006)\citenamefont {Adam},
  \citenamefont {H\'{e}ron},\ and\ \citenamefont {Laval}}]{Adam2006prl}%
  \BibitemOpen
  \bibfield  {author} {\bibinfo {author} {\bibfnamefont {J.}~\bibnamefont
  {Adam}}, \bibinfo {author} {\bibfnamefont {A.}~\bibnamefont {H\'{e}ron}},\
  and\ \bibinfo {author} {\bibfnamefont {G.}~\bibnamefont {Laval}},\ }\bibfield
   {title} {\emph {\bibinfo {title} {Dispersion and Transport of Energetic
  Particles due to the Interaction of Intense Laser Pulses with Overdense
  Plasmas}},\ }\href {https://doi.org/10.1103/PhysRevLett.97.205006} {\bibfield
   {journal} {\bibinfo  {journal} {Phys. Rev. Lett.}\ }\textbf {\bibinfo
  {volume} {97}},\ \bibinfo {pages} {205006} (\bibinfo {year}
  {2006})}\BibitemShut {NoStop}%
\bibitem [{\citenamefont {Green}\ \emph {et~al.}(2008)\citenamefont {Green},
  \citenamefont {Ovchinnikov}, \citenamefont {Evans}, \citenamefont {Akli},
  \citenamefont {Azechi}, \citenamefont {Beg}, \citenamefont {Bellei},
  \citenamefont {Freeman}, \citenamefont {Habara}, \citenamefont {Heathcote},
  \citenamefont {Key}, \citenamefont {King}, \citenamefont {Lancaster},
  \citenamefont {Lopes}, \citenamefont {Ma}, \citenamefont {MacKinnon},
  \citenamefont {Markey}, \citenamefont {McPhee}, \citenamefont {Najmudin},
  \citenamefont {Nilson}, \citenamefont {Onofrei}, \citenamefont {Stephens},
  \citenamefont {Takeda}, \citenamefont {Tanaka}, \citenamefont {Theobald},
  \citenamefont {Tanimoto}, \citenamefont {Waugh}, \citenamefont {Woerkom},
  \citenamefont {Woolsey}, \citenamefont {Zepf}, \citenamefont {Davies},\ and\
  \citenamefont {Norreys}}]{green2008prl}%
  \BibitemOpen
  \bibfield  {author} {\bibinfo {author} {\bibfnamefont {J.~S.}\ \bibnamefont
  {Green}}, \bibinfo {author} {\bibfnamefont {V.~M.}\ \bibnamefont
  {Ovchinnikov}}, \bibinfo {author} {\bibfnamefont {R.~G.}\ \bibnamefont
  {Evans}}, \bibinfo {author} {\bibfnamefont {K.~U.}\ \bibnamefont {Akli}},
  \bibinfo {author} {\bibfnamefont {H.}~\bibnamefont {Azechi}}, \bibinfo
  {author} {\bibfnamefont {F.~N.}\ \bibnamefont {Beg}}, \bibinfo {author}
  {\bibfnamefont {C.}~\bibnamefont {Bellei}}, \bibinfo {author} {\bibfnamefont
  {R.~R.}\ \bibnamefont {Freeman}}, \bibinfo {author} {\bibfnamefont
  {H.}~\bibnamefont {Habara}}, \bibinfo {author} {\bibfnamefont
  {R.}~\bibnamefont {Heathcote}}, \bibinfo {author} {\bibfnamefont {M.~H.}\
  \bibnamefont {Key}}, \bibinfo {author} {\bibfnamefont {J.~A.}\ \bibnamefont
  {King}}, \bibinfo {author} {\bibfnamefont {K.~L.}\ \bibnamefont {Lancaster}},
  \bibinfo {author} {\bibfnamefont {N.~C.}\ \bibnamefont {Lopes}}, \bibinfo
  {author} {\bibfnamefont {T.}~\bibnamefont {Ma}}, \bibinfo {author}
  {\bibfnamefont {A.~J.}\ \bibnamefont {MacKinnon}}, \bibinfo {author}
  {\bibfnamefont {K.}~\bibnamefont {Markey}}, \bibinfo {author} {\bibfnamefont
  {A.}~\bibnamefont {McPhee}}, \bibinfo {author} {\bibfnamefont
  {Z.}~\bibnamefont {Najmudin}}, \bibinfo {author} {\bibfnamefont
  {P.}~\bibnamefont {Nilson}}, \bibinfo {author} {\bibfnamefont
  {R.}~\bibnamefont {Onofrei}}, \bibinfo {author} {\bibfnamefont
  {R.}~\bibnamefont {Stephens}}, \bibinfo {author} {\bibfnamefont
  {K.}~\bibnamefont {Takeda}}, \bibinfo {author} {\bibfnamefont {K.~A.}\
  \bibnamefont {Tanaka}}, \bibinfo {author} {\bibfnamefont {W.}~\bibnamefont
  {Theobald}}, \bibinfo {author} {\bibfnamefont {T.}~\bibnamefont {Tanimoto}},
  \bibinfo {author} {\bibfnamefont {J.}~\bibnamefont {Waugh}}, \bibinfo
  {author} {\bibfnamefont {L.~V.}\ \bibnamefont {Woerkom}}, \bibinfo {author}
  {\bibfnamefont {N.~C.}\ \bibnamefont {Woolsey}}, \bibinfo {author}
  {\bibfnamefont {M.}~\bibnamefont {Zepf}}, \bibinfo {author} {\bibfnamefont
  {J.~R.}\ \bibnamefont {Davies}},\ and\ \bibinfo {author} {\bibfnamefont
  {P.~A.}\ \bibnamefont {Norreys}},\ }\bibfield  {title} {\emph {\bibinfo
  {title} {Effect of Laser Intensity on Fast-Electron-Beam Divergence in
  Solid-Density Plasmas}},\ }\href
  {https://doi.org/10.1103/PhysRevLett.100.039902} {\bibfield  {journal}
  {\bibinfo  {journal} {Phys. Rev. Lett.}\ }\textbf {\bibinfo {volume} {100}},\
  \bibinfo {pages} {039902} (\bibinfo {year} {2008})}\BibitemShut {NoStop}%
\bibitem [{\citenamefont {Fuchs}\ \emph {et~al.}(2007)\citenamefont {Fuchs},
  \citenamefont {Cecchetti}, \citenamefont {Borghesi}, \citenamefont
  {Grismayer}, \citenamefont {d’Humières}, \citenamefont {Antici},
  \citenamefont {Atzeni}, \citenamefont {Mora}, \citenamefont {Pipahl},
  \citenamefont {Romagnani}, \citenamefont {Schiavi}, \citenamefont {Sentoku},
  \citenamefont {Toncian}, \citenamefont {Audebert},\ and\ \citenamefont
  {Willi}}]{Fuchs2006prl}%
  \BibitemOpen
  \bibfield  {author} {\bibinfo {author} {\bibfnamefont {J.}~\bibnamefont
  {Fuchs}}, \bibinfo {author} {\bibfnamefont {C.~A.}\ \bibnamefont
  {Cecchetti}}, \bibinfo {author} {\bibfnamefont {M.}~\bibnamefont {Borghesi}},
  \bibinfo {author} {\bibfnamefont {T.}~\bibnamefont {Grismayer}}, \bibinfo
  {author} {\bibfnamefont {E.}~\bibnamefont {d’Humières}}, \bibinfo {author}
  {\bibfnamefont {P.}~\bibnamefont {Antici}}, \bibinfo {author} {\bibfnamefont
  {S.}~\bibnamefont {Atzeni}}, \bibinfo {author} {\bibfnamefont
  {P.}~\bibnamefont {Mora}}, \bibinfo {author} {\bibfnamefont {A.}~\bibnamefont
  {Pipahl}}, \bibinfo {author} {\bibfnamefont {L.}~\bibnamefont {Romagnani}},
  \bibinfo {author} {\bibfnamefont {A.}~\bibnamefont {Schiavi}}, \bibinfo
  {author} {\bibfnamefont {Y.}~\bibnamefont {Sentoku}}, \bibinfo {author}
  {\bibfnamefont {T.}~\bibnamefont {Toncian}}, \bibinfo {author} {\bibfnamefont
  {P.}~\bibnamefont {Audebert}},\ and\ \bibinfo {author} {\bibfnamefont
  {O.}~\bibnamefont {Willi}},\ }\bibfield  {title} {\emph {\bibinfo {title}
  {Laser-Foil Acceleration of High-Energy Protons in Small-Scale Plasma
  Gradients}},\ }\href {https://doi.org/10.1103/PhysRevLett.99.015002}
  {\bibfield  {journal} {\bibinfo  {journal} {Phys. Rev. Lett.}\ }\textbf
  {\bibinfo {volume} {99}},\ \bibinfo {pages} {015002} (\bibinfo {year}
  {2007})}\BibitemShut {NoStop}%
\bibitem [{\citenamefont {Wang}\ \emph {et~al.}(2021)\citenamefont {Wang},
  \citenamefont {Qi}, \citenamefont {Pan}, \citenamefont {Kong}, \citenamefont
  {Shou}, \citenamefont {Liu}, \citenamefont {Cao}, \citenamefont {Mei},
  \citenamefont {Xu}, \citenamefont {Liu},\ and\ \citenamefont
  {et~al.}}]{Wang2021}%
  \BibitemOpen
  \bibfield  {author} {\bibinfo {author} {\bibfnamefont {P.}~\bibnamefont
  {Wang}}, \bibinfo {author} {\bibfnamefont {G.}~\bibnamefont {Qi}}, \bibinfo
  {author} {\bibfnamefont {Z.}~\bibnamefont {Pan}}, \bibinfo {author}
  {\bibfnamefont {D.}~\bibnamefont {Kong}}, \bibinfo {author} {\bibfnamefont
  {Y.}~\bibnamefont {Shou}}, \bibinfo {author} {\bibfnamefont {J.}~\bibnamefont
  {Liu}}, \bibinfo {author} {\bibfnamefont {Z.}~\bibnamefont {Cao}}, \bibinfo
  {author} {\bibfnamefont {Z.}~\bibnamefont {Mei}}, \bibinfo {author}
  {\bibfnamefont {S.}~\bibnamefont {Xu}}, \bibinfo {author} {\bibfnamefont
  {Z.}~\bibnamefont {Liu}},\ and\ \bibinfo {author} {\bibnamefont {et~al.}},\
  }\bibfield  {title} {\emph {\bibinfo {title} {Fabrication of large-area
  uniform carbon nanotube foams as near-critical-density targets for
  laser–plasma experiments}},\ }\href {https://doi.org/10.1017/hpl.2021.18}
  {\bibfield  {journal} {\bibinfo  {journal} {High Power Laser Science and
  Engineering}\ }\textbf {\bibinfo {volume} {9}},\ \bibinfo {pages} {e29}
  (\bibinfo {year} {2021})}\BibitemShut {NoStop}%
\bibitem [{\citenamefont {Chen}\ \emph {et~al.}(2016)\citenamefont {Chen},
  \citenamefont {Gauthier}, \citenamefont {Bazalova-Carter}, \citenamefont
  {Bolanos}, \citenamefont {Glenzer}, \citenamefont {Riquier}, \citenamefont
  {Revet}, \citenamefont {Antici}, \citenamefont {Morabito}, \citenamefont
  {Propp}, \citenamefont {Starodubtsev},\ and\ \citenamefont
  {Fuchs}}]{chen2016rsi}%
  \BibitemOpen
  \bibfield  {author} {\bibinfo {author} {\bibfnamefont {S.~N.}\ \bibnamefont
  {Chen}}, \bibinfo {author} {\bibfnamefont {M.}~\bibnamefont {Gauthier}},
  \bibinfo {author} {\bibfnamefont {M.}~\bibnamefont {Bazalova-Carter}},
  \bibinfo {author} {\bibfnamefont {S.}~\bibnamefont {Bolanos}}, \bibinfo
  {author} {\bibfnamefont {S.}~\bibnamefont {Glenzer}}, \bibinfo {author}
  {\bibfnamefont {R.}~\bibnamefont {Riquier}}, \bibinfo {author} {\bibfnamefont
  {G.}~\bibnamefont {Revet}}, \bibinfo {author} {\bibfnamefont
  {P.}~\bibnamefont {Antici}}, \bibinfo {author} {\bibfnamefont
  {A.}~\bibnamefont {Morabito}}, \bibinfo {author} {\bibfnamefont
  {A.}~\bibnamefont {Propp}}, \bibinfo {author} {\bibfnamefont
  {M.}~\bibnamefont {Starodubtsev}},\ and\ \bibinfo {author} {\bibfnamefont
  {J.}~\bibnamefont {Fuchs}},\ }\bibfield  {title} {\emph {\bibinfo {title}
  {Absolute dosimetric characterization of Gafchromic EBT3 and HDv2 films using
  commercial flat-bed scanners and evaluation of the scanner response function
  variability}},\ }\href {https://doi.org/10.1063/1.4954921} {\bibfield
  {journal} {\bibinfo  {journal} {Review of Scientific Instruments}\ }\textbf
  {\bibinfo {volume} {87}},\ \bibinfo {pages} {073301} (\bibinfo {year}
  {2016})}\BibitemShut {NoStop}%
\bibitem [{\citenamefont {Lelasseux}\ and\ \citenamefont
  {Fuchs}(2020)}]{Lelasseux2020}%
  \BibitemOpen
  \bibfield  {author} {\bibinfo {author} {\bibfnamefont {V.}~\bibnamefont
  {Lelasseux}}\ and\ \bibinfo {author} {\bibfnamefont {J.}~\bibnamefont
  {Fuchs}},\ }\bibfield  {title} {\emph {\bibinfo {title} {Modelling energy
  deposition in TR image plate detectors for various ion types}},\ }\href@noop
  {} {\bibfield  {journal} {\bibinfo  {journal} {Journal of Instrumentation}\
  }\textbf {\bibinfo {volume} {15}}\bibinfo  {number} { (04)},\ \bibinfo
  {pages} {P04002}}\BibitemShut {NoStop}%
\bibitem [{\citenamefont {Ogura}\ \emph {et~al.}(2012)\citenamefont {Ogura},
  \citenamefont {Nishiuchi}, \citenamefont {Pirozhkov}, \citenamefont
  {Tanimoto}, \citenamefont {Sagisaka}, \citenamefont {Esirkepov},
  \citenamefont {Kando}, \citenamefont {Shizuma}, \citenamefont {Hayakawa},
  \citenamefont {Kiriyama}, \citenamefont {Shimomura}, \citenamefont {Kondo},
  \citenamefont {Kanazawa}, \citenamefont {Nakai}, \citenamefont {Sasao},
  \citenamefont {Sasao}, \citenamefont {Fukuda}, \citenamefont {Sakaki},
  \citenamefont {Kanasaki}, \citenamefont {Yogo}, \citenamefont {Bulanov},
  \citenamefont {Bolton},\ and\ \citenamefont {Kondo}}]{Ogura:12}%
  \BibitemOpen
\bibfield  {number} {  }\bibfield  {author} {\bibinfo {author} {\bibfnamefont
  {K.}~\bibnamefont {Ogura}}, \bibinfo {author} {\bibfnamefont
  {M.}~\bibnamefont {Nishiuchi}}, \bibinfo {author} {\bibfnamefont {A.~S.}\
  \bibnamefont {Pirozhkov}}, \bibinfo {author} {\bibfnamefont {T.}~\bibnamefont
  {Tanimoto}}, \bibinfo {author} {\bibfnamefont {A.}~\bibnamefont {Sagisaka}},
  \bibinfo {author} {\bibfnamefont {T.~Z.}\ \bibnamefont {Esirkepov}}, \bibinfo
  {author} {\bibfnamefont {M.}~\bibnamefont {Kando}}, \bibinfo {author}
  {\bibfnamefont {T.}~\bibnamefont {Shizuma}}, \bibinfo {author} {\bibfnamefont
  {T.}~\bibnamefont {Hayakawa}}, \bibinfo {author} {\bibfnamefont
  {H.}~\bibnamefont {Kiriyama}}, \bibinfo {author} {\bibfnamefont
  {T.}~\bibnamefont {Shimomura}}, \bibinfo {author} {\bibfnamefont
  {S.}~\bibnamefont {Kondo}}, \bibinfo {author} {\bibfnamefont
  {S.}~\bibnamefont {Kanazawa}}, \bibinfo {author} {\bibfnamefont
  {Y.}~\bibnamefont {Nakai}}, \bibinfo {author} {\bibfnamefont
  {H.}~\bibnamefont {Sasao}}, \bibinfo {author} {\bibfnamefont
  {F.}~\bibnamefont {Sasao}}, \bibinfo {author} {\bibfnamefont
  {Y.}~\bibnamefont {Fukuda}}, \bibinfo {author} {\bibfnamefont
  {H.}~\bibnamefont {Sakaki}}, \bibinfo {author} {\bibfnamefont
  {M.}~\bibnamefont {Kanasaki}}, \bibinfo {author} {\bibfnamefont
  {A.}~\bibnamefont {Yogo}}, \bibinfo {author} {\bibfnamefont {S.~V.}\
  \bibnamefont {Bulanov}}, \bibinfo {author} {\bibfnamefont {P.~R.}\
  \bibnamefont {Bolton}},\ and\ \bibinfo {author} {\bibfnamefont
  {K.}~\bibnamefont {Kondo}},\ }\bibfield  {title} {\emph {\bibinfo {title}
  {Proton acceleration to 40 MeV using a high intensity, high contrast optical
  parametric chirped-pulse amplification Ti\:sapphire hybrid laser system}},\
  }\href {https://doi.org/10.1364/OL.37.002868} {\bibfield  {journal} {\bibinfo
   {journal} {Opt. Lett.}\ }\textbf {\bibinfo {volume} {37}},\ \bibinfo {pages}
  {2868--2870} (\bibinfo {year} {2012})}\BibitemShut {NoStop}%
\bibitem [{\citenamefont {Wagner}\ \emph
  {et~al.}(2016{\natexlab{b}})\citenamefont {Wagner}, \citenamefont {Deppert},
  \citenamefont {Brabetz}, \citenamefont {Fiala}, \citenamefont {Kleinschmidt},
  \citenamefont {Poth}, \citenamefont {Schanz}, \citenamefont {Tebartz},
  \citenamefont {Zielbauer}, \citenamefont {Roth}, \citenamefont {St\"ohlker},\
  and\ \citenamefont {Bagnoud}}]{wagner2016prl}%
  \BibitemOpen
  \bibfield  {author} {\bibinfo {author} {\bibfnamefont {F.}~\bibnamefont
  {Wagner}}, \bibinfo {author} {\bibfnamefont {O.}~\bibnamefont {Deppert}},
  \bibinfo {author} {\bibfnamefont {C.}~\bibnamefont {Brabetz}}, \bibinfo
  {author} {\bibfnamefont {P.}~\bibnamefont {Fiala}}, \bibinfo {author}
  {\bibfnamefont {A.}~\bibnamefont {Kleinschmidt}}, \bibinfo {author}
  {\bibfnamefont {P.}~\bibnamefont {Poth}}, \bibinfo {author} {\bibfnamefont
  {V.~A.}\ \bibnamefont {Schanz}}, \bibinfo {author} {\bibfnamefont
  {A.}~\bibnamefont {Tebartz}}, \bibinfo {author} {\bibfnamefont
  {B.}~\bibnamefont {Zielbauer}}, \bibinfo {author} {\bibfnamefont
  {M.}~\bibnamefont {Roth}}, \bibinfo {author} {\bibfnamefont {T.~.}\
  \bibnamefont {St\"ohlker}},\ and\ \bibinfo {author} {\bibfnamefont
  {V.}~\bibnamefont {Bagnoud}},\ }\bibfield  {title} {\emph {\bibinfo {title}
  {Maximum Proton Energy above 85 MeV from the Relativistic Interaction of
  Laser Pulses with Micrometer Thick CH2 Targets}},\ }\href
  {https://doi.org/10.1103/PhysRevLett.116.205002} {\bibfield  {journal}
  {\bibinfo  {journal} {Phys. Rev. Lett.}\ }\textbf {\bibinfo {volume} {116}},\
  \bibinfo {pages} {205002} (\bibinfo {year} {2016}{\natexlab{b}})}\BibitemShut
  {NoStop}%
\bibitem [{\citenamefont {Chen}\ \emph {et~al.}(2008)\citenamefont {Chen},
  \citenamefont {Back}, \citenamefont {Bartal}, \citenamefont {Beg},
  \citenamefont {Eder}, \citenamefont {Link}, \citenamefont {MacPhee},
  \citenamefont {Ping}, \citenamefont {Song}, \citenamefont {Throop},\ and\
  \citenamefont {Van~Woerkom}}]{chen2008rsi}%
  \BibitemOpen
  \bibfield  {author} {\bibinfo {author} {\bibfnamefont {H.}~\bibnamefont
  {Chen}}, \bibinfo {author} {\bibfnamefont {N.~L.}\ \bibnamefont {Back}},
  \bibinfo {author} {\bibfnamefont {T.}~\bibnamefont {Bartal}}, \bibinfo
  {author} {\bibfnamefont {F.~N.}\ \bibnamefont {Beg}}, \bibinfo {author}
  {\bibfnamefont {D.~C.}\ \bibnamefont {Eder}}, \bibinfo {author}
  {\bibfnamefont {A.~J.}\ \bibnamefont {Link}}, \bibinfo {author}
  {\bibfnamefont {A.~G.}\ \bibnamefont {MacPhee}}, \bibinfo {author}
  {\bibfnamefont {Y.}~\bibnamefont {Ping}}, \bibinfo {author} {\bibfnamefont
  {P.~M.}\ \bibnamefont {Song}}, \bibinfo {author} {\bibfnamefont
  {A.}~\bibnamefont {Throop}},\ and\ \bibinfo {author} {\bibfnamefont
  {L.}~\bibnamefont {Van~Woerkom}},\ }\bibfield  {title} {\emph {\bibinfo
  {title} {Absolute calibration of image plates for electrons at energy between
  100keV and 4MeV}},\ }\href {https://doi.org/10.1063/1.2885045} {\bibfield
  {journal} {\bibinfo  {journal} {Review of Scientific Instruments}\ }\textbf
  {\bibinfo {volume} {79}},\ \bibinfo {pages} {033301} (\bibinfo {year}
  {2008})}\BibitemShut {NoStop}%
\bibitem [{\citenamefont {Alejo}\ \emph {et~al.}(2015)\citenamefont {Alejo},
  \citenamefont {Ahmed}, \citenamefont {Green}, \citenamefont {Mirfayzi},
  \citenamefont {Borghesi},\ and\ \citenamefont {Kar}}]{Alejo2015}%
  \BibitemOpen
  \bibfield  {author} {\bibinfo {author} {\bibfnamefont {A.}~\bibnamefont
  {Alejo}}, \bibinfo {author} {\bibfnamefont {H.}~\bibnamefont {Ahmed}},
  \bibinfo {author} {\bibfnamefont {A.}~\bibnamefont {Green}}, \bibinfo
  {author} {\bibfnamefont {S.~R.}\ \bibnamefont {Mirfayzi}}, \bibinfo {author}
  {\bibfnamefont {M.}~\bibnamefont {Borghesi}},\ and\ \bibinfo {author}
  {\bibfnamefont {S.}~\bibnamefont {Kar}},\ }\bibfield  {title} {\emph
  {\bibinfo {title} {Recent advances in laser-driven neutron sources}},\ }\href
  {https://doi.org/10.1393/ncc/i2015-15188-8} {\bibfield  {journal} {\bibinfo
  {journal} {Il Nuovo Cimento}\ }\textbf {\bibinfo {volume} {38}},\ \bibinfo
  {pages} {188} (\bibinfo {year} {2015})}\BibitemShut {NoStop}%
\bibitem [{\citenamefont {Mirfayzi}\ \emph {et~al.}(2015)\citenamefont
  {Mirfayzi}, \citenamefont {Kar}, \citenamefont {Ahmed}, \citenamefont
  {Krygier}, \citenamefont {Green}, \citenamefont {Alejo}, \citenamefont
  {Clarke}, \citenamefont {Freeman}, \citenamefont {Fuchs}, \citenamefont
  {Jung}, \citenamefont {Kleinschmidt}, \citenamefont {Morrison}, \citenamefont
  {Najmudin}, \citenamefont {Nakamura}, \citenamefont {Norreys}, \citenamefont
  {Oliver}, \citenamefont {Roth}, \citenamefont {Vassura}, \citenamefont
  {Zepf},\ and\ \citenamefont {Borghesi}}]{Mirfayzi2015rsi}%
  \BibitemOpen
  \bibfield  {author} {\bibinfo {author} {\bibfnamefont {S.~R.}\ \bibnamefont
  {Mirfayzi}}, \bibinfo {author} {\bibfnamefont {S.}~\bibnamefont {Kar}},
  \bibinfo {author} {\bibfnamefont {H.}~\bibnamefont {Ahmed}}, \bibinfo
  {author} {\bibfnamefont {A.~G.}\ \bibnamefont {Krygier}}, \bibinfo {author}
  {\bibfnamefont {A.}~\bibnamefont {Green}}, \bibinfo {author} {\bibfnamefont
  {A.}~\bibnamefont {Alejo}}, \bibinfo {author} {\bibfnamefont
  {R.}~\bibnamefont {Clarke}}, \bibinfo {author} {\bibfnamefont {R.~R.}\
  \bibnamefont {Freeman}}, \bibinfo {author} {\bibfnamefont {J.}~\bibnamefont
  {Fuchs}}, \bibinfo {author} {\bibfnamefont {D.}~\bibnamefont {Jung}},
  \bibinfo {author} {\bibfnamefont {A.}~\bibnamefont {Kleinschmidt}}, \bibinfo
  {author} {\bibfnamefont {J.~T.}\ \bibnamefont {Morrison}}, \bibinfo {author}
  {\bibfnamefont {Z.}~\bibnamefont {Najmudin}}, \bibinfo {author}
  {\bibfnamefont {H.}~\bibnamefont {Nakamura}}, \bibinfo {author}
  {\bibfnamefont {P.}~\bibnamefont {Norreys}}, \bibinfo {author} {\bibfnamefont
  {M.}~\bibnamefont {Oliver}}, \bibinfo {author} {\bibfnamefont
  {M.}~\bibnamefont {Roth}}, \bibinfo {author} {\bibfnamefont {L.}~\bibnamefont
  {Vassura}}, \bibinfo {author} {\bibfnamefont {M.}~\bibnamefont {Zepf}},\ and\
  \bibinfo {author} {\bibfnamefont {M.}~\bibnamefont {Borghesi}},\ }\bibfield
  {title} {\emph {\bibinfo {title} {Calibration of time of flight detectors
  using laser-driven neutron source}},\ }\href
  {https://doi.org/10.1063/1.4923088} {\bibfield  {journal} {\bibinfo
  {journal} {Review of Scientific Instruments}\ }\textbf {\bibinfo {volume}
  {86}},\ \bibinfo {pages} {073308} (\bibinfo {year} {2015})}\BibitemShut
  {NoStop}%
\bibitem [{DDO()}]{DDOT}%
  \BibitemOpen
  \href
  {https://www.prodyntech.com/wp-content/uploads/2013/09/D-Dot-Free-Field-W2.pdf}
  {\bibinfo {title} {Electric Field Sensors, D-Dot free field-radial output
  (R)}},\ \bibinfo {howpublished} {Data Sheet Prodyn Technologies},\ \bibinfo
  {note} {accessed on the 22th July 2021}\BibitemShut {NoStop}%
\bibitem [{BDO()}]{BDOT}%
  \BibitemOpen
  \href
  {https://prodyntech.com/wp-content/uploads/2020/12/RB-230-DATA-SHEET-05272018193141.pdf}
  {\bibinfo {title} {Radiation hardened B-Dot Sensor - Model RB-230(R)}},\
  \bibinfo {howpublished} {Data Sheet Prodyn Technologies},\ \bibinfo {note}
  {accessed on the 22th July 2021}\BibitemShut {NoStop}%
\bibitem [{BAL()}]{BALUN}%
  \BibitemOpen
  \href
  {https://www.prodyntech.com/wp-content/uploads/2013/09/Balun-Spec-Sheet-01292018212240.pdf}
  {\bibinfo {title} {Prodyn Baluns}},\ \bibinfo {howpublished} {Data Sheet
  Prodyn Technologies},\ \bibinfo {note} {accessed on the 22th July
  2021}\BibitemShut {NoStop}%
\bibitem [{\citenamefont {Thaury}\ \emph {et~al.}(2007)\citenamefont {Thaury},
  \citenamefont {Qu{\'e}r{\'e}}, \citenamefont {Geindre}, \citenamefont {Levy},
  \citenamefont {Ceccotti}, \citenamefont {Monot}, \citenamefont {Bougeard},
  \citenamefont {R{\'e}au}, \citenamefont {d'Oliveira}, \citenamefont
  {Audebert}, \citenamefont {Marjoribanks},\ and\ \citenamefont
  {Martin}}]{Thaury2007natphys}%
  \BibitemOpen
  \bibfield  {author} {\bibinfo {author} {\bibfnamefont {C.}~\bibnamefont
  {Thaury}}, \bibinfo {author} {\bibfnamefont {F.}~\bibnamefont
  {Qu{\'e}r{\'e}}}, \bibinfo {author} {\bibfnamefont {J.-P.}\ \bibnamefont
  {Geindre}}, \bibinfo {author} {\bibfnamefont {A.}~\bibnamefont {Levy}},
  \bibinfo {author} {\bibfnamefont {T.}~\bibnamefont {Ceccotti}}, \bibinfo
  {author} {\bibfnamefont {P.}~\bibnamefont {Monot}}, \bibinfo {author}
  {\bibfnamefont {M.}~\bibnamefont {Bougeard}}, \bibinfo {author}
  {\bibfnamefont {F.}~\bibnamefont {R{\'e}au}}, \bibinfo {author}
  {\bibfnamefont {P.}~\bibnamefont {d'Oliveira}}, \bibinfo {author}
  {\bibfnamefont {P.}~\bibnamefont {Audebert}}, \bibinfo {author}
  {\bibfnamefont {R.}~\bibnamefont {Marjoribanks}},\ and\ \bibinfo {author}
  {\bibfnamefont {P.}~\bibnamefont {Martin}},\ }\bibfield  {title} {\emph
  {\bibinfo {title} {Plasma mirrors for ultrahigh-intensity optics}},\ }\href
  {https://doi.org/10.1038/nphys595} {\bibfield  {journal} {\bibinfo  {journal}
  {Nature Physics}\ }\textbf {\bibinfo {volume} {3}},\ \bibinfo {pages}
  {424--429} (\bibinfo {year} {2007})}\BibitemShut {NoStop}%
\end{thebibliography}%

% \appendix*
% \section*{APPENDIX: DETAILED CHARACTERIZATION TABLE}

\end{document}